# ON PERTURBATION THEORY FOR THE STURM-LIOUVILLE PROBLEM WITH VARIABLE COEFFICIENTS, PART II

## Vladimir Kalitvianski

vladimir.kalitvianski@wanadoo.fr
Grenoble, 2018

I study some possibilities of analytically solving a particular Sturm-Liouville problem with step-wise (piece-constant) coefficients with help of an iterative procedure mentioned in my previous paper. I construct simple, but very accurate analytical formulae for calculating the ground state eigenfunction and the eigenvalue as well as for calculating the first eigenfunction. I study numerical precision of the obtained approximations together with the perturbation theory results.

## INTRODUCTION

The Sturm-Liouville problem (SLP), understood here in a narrow sense of obtaining the eigenfunctions and eigenvalues, arises in many practical applications. Despite wide use of numerical approaches, the analytical solutions also represent a certain scientific value, especially if their physical sense is clear and the analytical formulae are simple. In this work I study possibilities of constructing analytically the SLP perturbative solutions and analyze their accuracy. They may be useful for quick estimations in heat conduction, diffusion, quantum mechanical problems, etc. The consideration is made on a "physical" level of rigor for simplicity.

This article is a follow up of my previous paper [1] where I proposed a new formula for the "ground state" eigenfunction $\psi_0$. In the present article I first develop this direction more and compare different approximate and exact solutions (**Sections 1-9**). As well, I test short analytical formulae for the first-order eigenfunctions $\psi_n^{PT(1)}(x)$ (**Sections 2, 8**), the comparison, which was missing in my previous publications [1], [2].

### 1. A perturbation theory formulation

Here I study the following particular SLP:

$$\left[\frac{d^2}{dx^2} + \lambda \cdot r(x)\right]\psi(x) = 0, \quad \psi(a) = 0, \quad \psi'(b) = 0. \tag{1}$$

By the variable changes [3]:

$$z(x) = z_a + \int_a^x \sqrt{r(x')}dx', \quad \tilde{r}(z) = r(x(z)), \quad \Phi(z) = \tilde{r}^{1/4}\psi(x(z)), \tag{2}$$

this problem can be transformed into a Schrödinger-like equation [2], [3]:

$$\left[\frac{d^2}{dz^2} + \lambda - U(z)\right]\Phi(z) = 0 \quad \text{with} \quad U(z) = \frac{\left(\tilde{r}^{1/4}\right)''}{\tilde{r}^{1/4}}. \tag{3}$$

These variable changes (2) look like those in a WKB approximation, but here there are no turning points since $r(x) > 0$ everywhere in our case. (The prime sign $(...)'$ means here a derivative with respect to the function argument.)

When the material properties $\tilde{r}(z)$ change smoothly (slowly with $z$), the derivatives of $\tilde{r}(z)$ in $U(z)$ are "small" and one can apply the perturbation theory (PT) to calculate the eigenfunctions and eigenvalues. I will not consider smooth $\tilde{r}(z)$ here. I will go directly to a case of a step-wise (or a piece-constant) function $\tilde{r}(z)$, namely:

$$r(x) = \begin{cases} r_1, & 0 \le x \le x_1 \\ r_2, & x_1 \le x \le 1 \end{cases} \quad \Rightarrow \quad \tilde{r}(z) = \begin{cases} r_1, & 0 \le z \le z_1, \quad z_1 = z(x_1) \\ r_2, & z_1 \le z \le z(1). \end{cases} \tag{4}$$

(The interval $(a,b)$ of $x$-variations is normalized below for simplicity to be $(0,1)$). We will consider $\tilde{r}(z)$ as a continuous, but extremely rapidly changing function at $z = z_1$ (a "two-layer" system). In a certain sense we may call (7) a "discontinuous function". As the matrix elements $U_{nm}$ for (4) diverge ($U(z)$ contains the Dirac's delta-function squared), we will not use them in our calculations.

Let us remember that an exact function $\psi_n(x)$ has different "*spatial frequencies*" and different *local amplitudes* in different layers, but it is an everywhere continuous function with a continuous derivative including the point $x = x_1$. The variable changes (2) catch well these properties in case of "smooth" functions $r(x)$ (it is well known from the WKB approximation applications, see also Fig. 1).

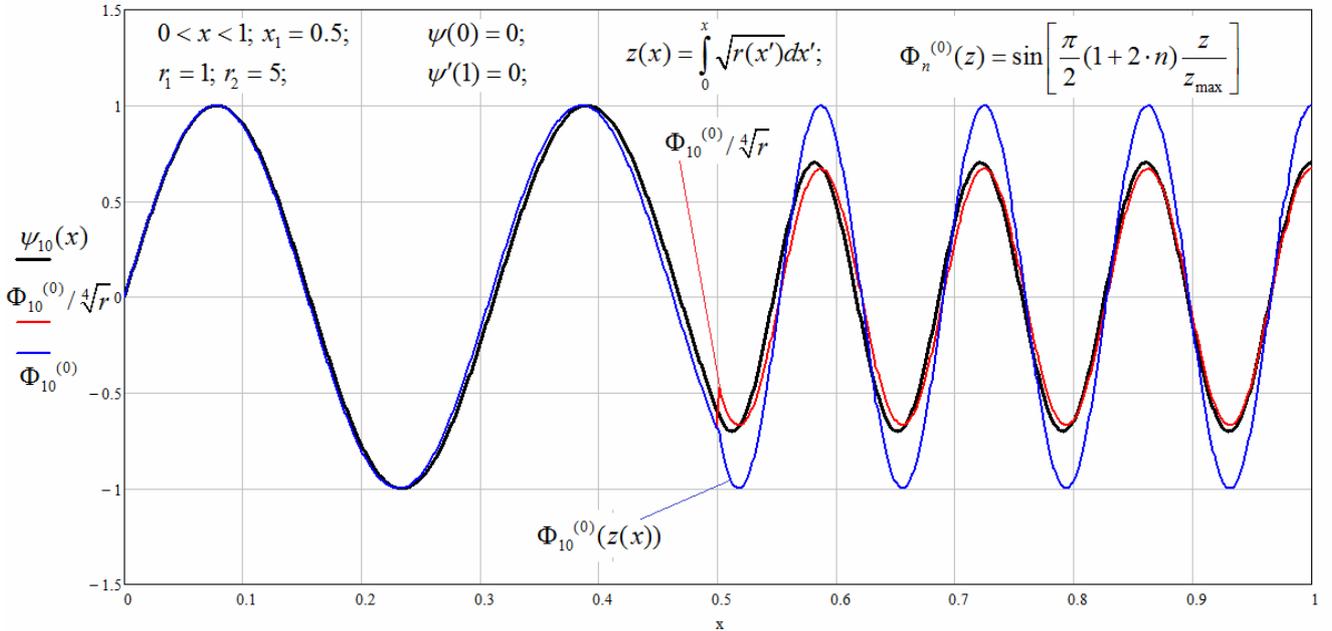

**Fig.1**

The exact eigenfunction $\psi_{n=10}(x)$ is an everywhere continuous function with a continuous derivative $\psi_{10}'(x)$ (the black line). The zeroth-order (or WKB-like) approximation to $\psi_{10}(x)$ (2) is discontinuous due to the factor $1/\sqrt[4]{r(x)}$ (the red line), but in an "integral sense" it mimics well the exact $\psi_{10}(x)$. The normalization factors here are chosen to reproduce the same amplitude as $\psi_{10}(x)$ within $0 < x < x_1$. The exact function $\psi_{10}(x)$ itself was not normalized in this figure, though.

## 2. Another Perturbation Theory formulation

We can make variable changes *without* this "discontinuous" factor at $\psi$, for example:

$$z(x) = \int_0^x \sqrt{r(x')}dx', \quad \tilde{r}(z) = r(x(z)), \quad \varphi(z) = \psi(x(z)). \tag{5}$$

Then we obtain the following exact equation:

$$\left[\frac{d^2}{dz^2} + \lambda - \hat{V}(z)\right]\varphi(z) = 0 \quad \text{with} \quad \hat{V}(z) = \left(\frac{1}{\sqrt{\tilde{r}}}\right)' \sqrt{\tilde{r}}\,\frac{d}{dz} = -\left(\ln\sqrt{\tilde{r}}\right)'\frac{d}{dz}. \tag{6}$$

If we treat $\hat{V}$ as a perturbation, then, figuratively speaking, we will take "blue lines" like that in Fig. 1 as the initial approximation $\varphi_n^{(0)}(z(x))$ to the "black ones" $\psi_n(x)$ (note, here $\varphi_n^{(0)}(z(x)) = \Phi_n^{(0)}(z(x))$).

The spatial "frequencies" and the local amplitudes are corrected in (6) exclusively with the finite perturbative terms now. Indeed, as far as the logarithm derivative in $\hat{V}(z)$ is proportional to the first degree of the Dirac's delta-function, the matrix elements are always *finite*. For (4) they are the following [1], [2]:

$$V_{nn} = \frac{1}{2}\varphi_n^{(0)}(z_1)\cdot\varphi_m^{(0)\prime}(z_1)\cdot\ln\left(\frac{r_1}{r_2}\right), \quad \lambda_n^{PT(0)} = \frac{\pi^2(2n+1)^2}{4z_b^2}; \; \varphi_0^{(0)}(z) = \sqrt{\frac{2}{z_b}}\sin\left(\sqrt{\lambda_n^{(0)}}\,z\right). \tag{7}$$

They are useful for eigenfunction and eigenvalue calculations, for example, in the problem (1):

$$\lambda_n^{PT(1)} = \lambda_n^{PT(0)} + V_{nn} = \frac{\pi^2(2n+1)^2}{4z_b^2} + \varepsilon\cdot\frac{\pi(2n+1)}{4z_b^2}\sin\left(\pi(2n+1)\frac{z_1}{z_b}\right); \; n = 0,1,2,...; \tag{8a}$$

$$\lambda_n^{PT(2)} = \lambda_n^{PT(1)} + \varepsilon^2\frac{\sqrt{\lambda_n^{(0)}}\sin\left(2\sqrt{\lambda_n^{(0)}}z_1\right)}{8z_b}\left\{\frac{\sin\left(2\sqrt{\lambda_n^{(0)}}z_1\right)}{2\sqrt{\lambda_n^{(0)}}z_b} - \left(1 - 2\frac{z_1}{z_b}\right)\cos\left(2\sqrt{\lambda_n^{(0)}}z_1\right)\right\}, \tag{8b}$$

where $\varepsilon = \ln\left(r_1/r_2\right)$, $z_1 = z(x_1)$, $z_b = z(1)$. For small and moderate values of $\varepsilon$ the second-order PT-approximations $\lambda_n^{PT(2)}$ are sufficient (see **Section 8** and **Appendix 1** for their derivation).

## 3. The lowest eigenvalue $\lambda_0$

The original equation (1) has a Green's function $G_\lambda(x,x')$ whose spectral representation is well known:

$$\frac{d^2}{dx^2}G_\lambda(x,y) + \lambda\cdot r(x)\cdot G_\lambda(x,y) = -\delta(x-y), \quad G_\lambda(x,y) = \sum_{n=0}^{\infty}\frac{\psi_n(x)\psi_n(y)}{\lambda_n - \lambda}. \tag{9}$$

Note, a "free" Green's function $G_{\lambda=0}(x,y) \equiv G_0(x,y)$ does not depend on $r(x)$ at all, so it can be expressed via exact $\psi_n$ and $\lambda_n$ of any particular problem. $G_0(x,y)$ is exactly constructible and in our case it is the following: $G_0(x,y) = x, \forall x \le y$, $G_0(x,y) = y, \forall x \ge y$. In other words, $G_0(x,y)$ does not depend on $y$ or on $x$ when $x \le y$ or $x \ge y$ correspondingly, $y$ being a parameter in it. Here is a picture of $G_0(x,0.5)$:

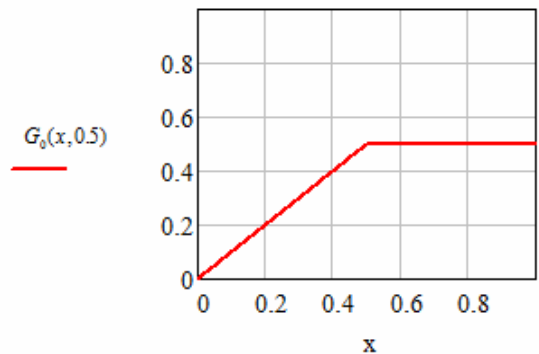

However some integral relationships involving $G_0(x,y)$ depend on $r(x)$ of a particular problem, for example, the sum rule:

$$\int_0^1 G_0(x,x)r(x)dx = \sum_{n=0}^\infty \frac{1}{\lambda_n}.$$ (10)

Basing on this sum rule I proposed a very simple and accurate analytical formula for the "ground state" eigenvalue $\lambda_0$:

$$\lambda_0^{GF(1)} = \left\{ \underbrace{\frac{\left(\lambda_0^{(0)}\right)^{-1}}{\frac{4z_b^2}{\pi^2}}}\left[1 - \frac{\varepsilon}{\pi}\sin\left(\pi\frac{z_1}{z_b}\right)\overbrace{\frac{\frac{4z_b^2}{\pi^2}V_{00}}{}}\right] + \underbrace{\frac{\int G_0(x,x)r(x)dx}{z_1(z_b - z_1)e^{-\varepsilon/2}}} - \underbrace{\frac{\int \Gamma_0^{PT(1)}(z,z)dz}{z_1(z_b - z_1)}\left(1 - \frac{\varepsilon}{2}\right)} \right\}^{-1}.$$ (11)

It has a numerical accuracy comparable with the third-order perturbation theory formula $\lambda_n^{PT(3)}$ [1], [2].

## 4. The "ground state" eigenfunction $\psi_0$

The "free" Green's function can also be used for calculation of the "ground state" eigenfunction $\psi_0(x)$. I have already tried such an approach in the previous paper. The idea was simple: as far as the WKB-like functions $\psi_n^{(0)}(x) = \Phi_n^{(0)}(z(x))/\sqrt[4]{r(x)}$ reproduce well the spatial frequencies and *amplitudes* of the exact ones $\psi_n(x)$ in an integral sense (see Fig. 1), then they might be "well orthogonal" to the exact ones. Namely, if $\psi_0^{(0)}(x)$ is "sufficiently orthogonal" to all $\psi_n(x)$ for $n \geq 1$, then an approximate $\psi_0^{GF}(x)$ can be constructed from the exact spectral representation of $G_0(x,y) = \frac{\psi_0(x)\psi_0(y)}{\lambda_0} + \sum_{n=1}^\infty \frac{\psi_n(x)\psi_n(y)}{\lambda_n}$, as well as from (1):

$$\frac{d^2}{dx^2}\psi_0(x) = \underbrace{-\lambda_0 r(x)\psi_0(x)}_{\text{A "source" term}} \Rightarrow \psi_0(x) = \lambda_0\left(\psi_0(y), G_0(x,y)\right),$$ (12)

where the bracket $\left(\phi(y), \eta(y)\right)$ denotes here a scalar product with weight $r(y)$. Thus, one can study the approximation with some $\psi_0^{(0)}$ on the right-hand side:

$$\psi_0(x)^{GF} \propto \int_0^1 r(y)\cdot\psi_0(y)^{(0)}\cdot G_0(x,y)dy.$$ (13)

If the contribution of the sum $\sum_{n=1}^\infty \psi_n(x)\frac{\left(\psi_0(y)^{(0)}, \psi_n(y)\right)}{\lambda_n}$ is negligible due to an effective orthogonality and rapidly growing denominators $\lambda_{n\geq 1}$, then integral (13) is just $\psi_0(x)$ with some coefficient. If so, then one may normalize the so obtained eigenfunction for further use and that's it. In fact, I tried to test the following formula:

$$\psi_0(x)^{GF} = \lambda_0^{GF(1)}\cdot\int_0^1 r(y)\cdot\psi_0(y)^{(0)}\cdot G_0(x,y)dy.$$ (14)

Here I used a very precise analytical approximation $\lambda_0^{GF(1)}$ (11) instead of unknown exact value $\lambda_0$ and I hoped that no additional normalization would be necessary because the exact sum rule (11) implies a normalized $\psi_0(x)$ on the left-hand side. The analytical expression for (14) with a WKB-like $\psi_0^{(0)}$ was simple, smooth, and it is the following:

$$\psi_0{}^{GF(0)}(x) = 4\lambda_0{}^{GF(1)}\frac{z(1)^2}{\pi^2}\Bigg[ \varphi_0{}^{(0)}(z(x))\left( \frac{\theta(x_1-x)}{\sqrt[4]{r_1}} + \frac{\theta(x-x_1)}{\sqrt[4]{r_2}} \right) +$$

$$+\left( \frac{1}{\sqrt[4]{r_1}} - \frac{1}{\sqrt[4]{r_2}} \right)\varphi_0{}^{(0)}(z(x_1))\cdot\theta(x-x_1) + \sqrt{\frac{2}{z_b}}\frac{\pi}{2z_b}\cos\left( \frac{\pi}{2}\frac{z(x_1)}{z_b} \right)G_0(x,x_1)\left( r_2{}^{1/4} - r_1{}^{1/4} \right) \Bigg]. \qquad (15)$$

Here $\theta(x)$ is the Heaviside stepwise $(0 \to 1)$ function. Numerical calculations with different $r_1$, $r_2$, and $x_1$ demonstrated a good accuracy of this formula. In particular, the figures below join two cases: ($r_1 = 1$, $r_2 = 4.5$) and ($r_1 = 4.5$, $r_2 = 1$), both with variable $0 \le x_1 \le 1$. Comparison was made with the exact normalized two-layer system solutions $\psi_0(x)$.

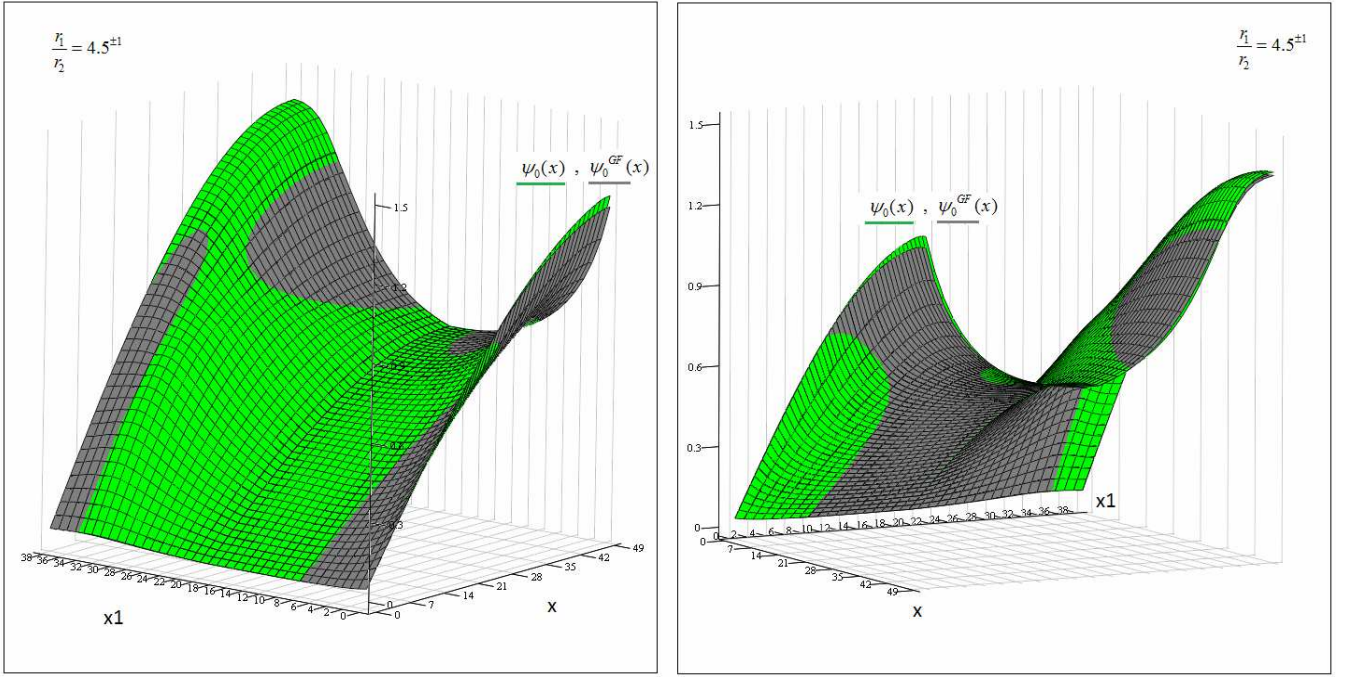

**Fig. 2.** Surfaces of $\psi_0(x)$ (green) and $\psi_0{}^{GF(0)}(x)$ (grey) for different values of $x_1$ ranging from 0.95 to 0.05 with a step of 0.05. The central line N°19 corresponds to the limiting case of a homogeneous one-layer system with $r = 4.5$, i.e., to the case ($r_1 = 1$, $r_2 = 4.5$, $x_1 = 0$) or to the case ($r_1 = 4.5$, $r_2 = 1$, $x_1 = 1$) (see Fig. 4 for details). The argument $x$ runs a discrete set of 50 values from 0.02 to 1 with a step $\Delta x = 0.02$ (realised as $X_i = i/50$, $i = 1, 2, ..., 50$). The points $x = 0$ where $\psi_0(0) = 0$ were not included in the resulting table in order to avoid the uncertainty $0/0$ in the "ratio" figures below. These surfaces generalize Fig. 12 and Fig. 13 from [1].

The table of obtained data, for example, for $\psi_0(x)$, represents a rectangular matrix or a two-dimensional array $(\psi_0)_{k,l}$. The graphic software I used to plot the surfaces has arrays that begin with index 0 by default. That is why in these figures the plots $(\psi_0)_{k,l}$ and $\left(\psi_0{}^{GF}\right)_{k,l}$ start from zeroes – the axes correspond to the array indices. The further relationships between $x_1$ and the corresponding index is explained in Fig. 4.

Apart from these figures, I plotted the ratios $\psi_0{}^{GF(0)}(x)/\psi_0(x)$. One can see in Fig. 3 that the "ratio surface" is close to unity.

When we take a "horizontal look" at it "from the back" of Fig. 3 (corresponding to different values of $x_1$), we obtain Fig. 4.

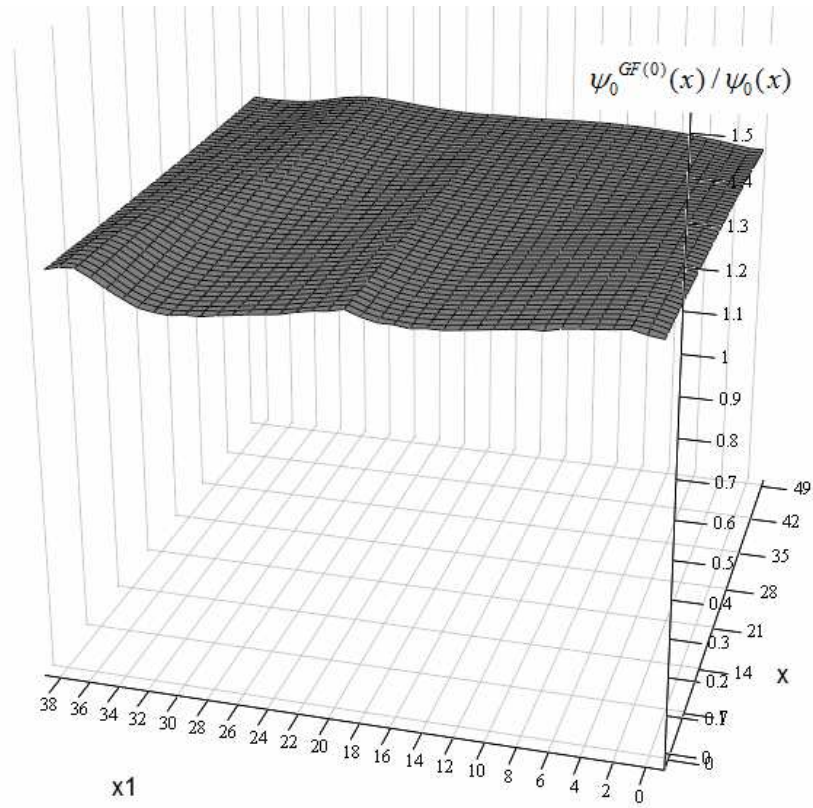

**Fig. 3.** The ratio $\psi_0^{GF(0)}(x)\,/\,\psi_0(x)$ from (15) (a picture with a perspective view).

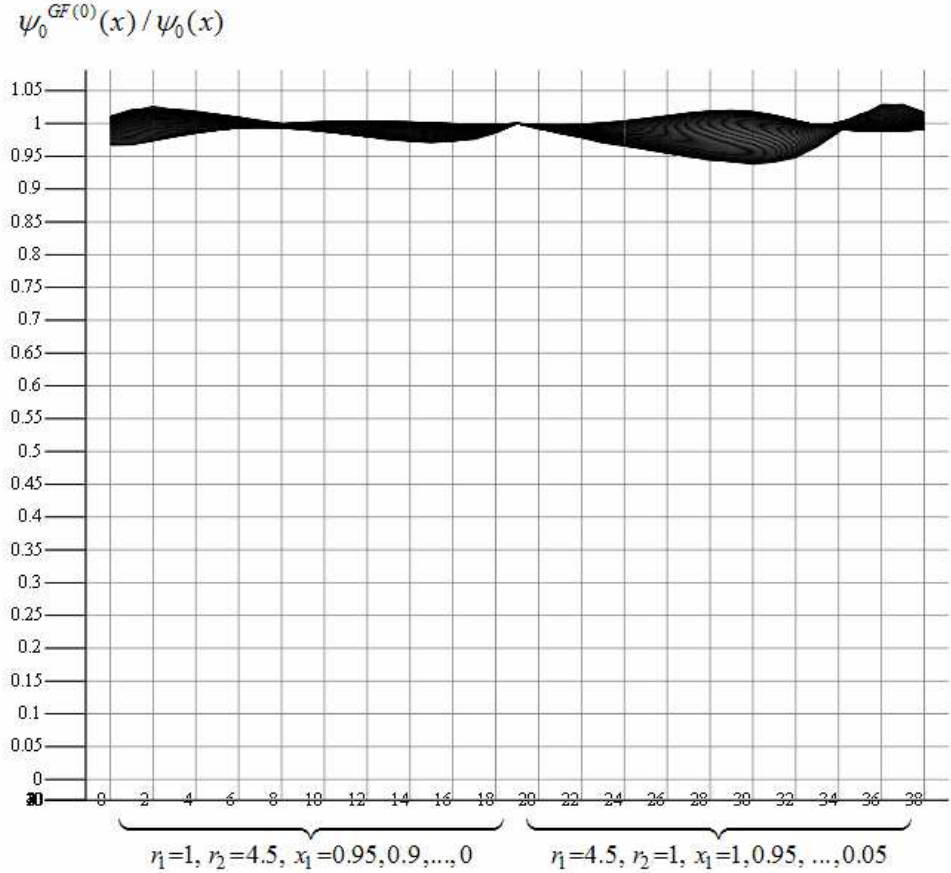

**Fig. 4.** The ratio profile $\psi_0^{GF(0)}(x)\,/\,\psi_0(x)$ from (15), a "horizontal projection" of Fig. 3 on the plane $\left(x_1, Z\right)$.

$\psi_0^{GF(0)}$ "winding" around the exact function $\psi_0$ in Fig. 2 (or in Fig. 13 from [1]) is seen here as the ratio "winding" around unity. As the ratio turns around unity nearly symmetrically, no additional normalization may improve the relative error "centring". Thus, formula (14) solves the problem with the relative accuracy $\leq \pm 5\%$.

These results are already contained in [1]: formula (23c), **Appendices 3** and **6**, etc.

In this paper I amend these results with new ones obtained on the way of exploiting the Green's function sum rule.

### 5. "No WKB" initial approximation

I noticed that for small differences between $r_1$ and $r_2$ the "jump" of $\psi_0^{(0)}(x)$ at $x = x_1$ is relatively small because a "square root of a square root" factor $\sqrt[4]{r}$ is not too sensitive to small variations of $r(x)$. Then I thought that this discontinuous factor might be avoided in the right-hand side of (13). In other words, I tried also $\phi_0^{(0)}(x) = \Phi_0^{(0)}(z(x)) = \varphi_0^{(0)}(z(x)) \propto \sin\left(\dfrac{\pi}{2}\dfrac{z(x)}{z_b}\right)$ as the initial approximation and I obtained:

$$\psi_0^{GF(0)}(x) \propto \left[ \sin\left(\frac{\pi}{2}\frac{z(x)}{z_b}\right) + \frac{\pi}{2z_b}\cos\left(\frac{\pi}{2}\frac{z(x_1)}{z_b}\right)G_0(x,x_1)\left(\sqrt{r_2}-\sqrt{r_1}\right) \right]. \tag{16}$$

Of course, the trial function $\phi_0^{(0)}(x)$ needs an additional "normalization factor" in order to coincide with $\psi_0(x)$ when $r(x) = const \neq 1$. First I used $\left\langle\sqrt{r}\right\rangle$ for $\varphi_0^{(0)}(z(x))$ (see **Section 6** in [1]) and found that the ratio (like that in Fig. 4) was not centred around unity for our $r_1$ and $r_2$, so this choice was not as perfect as I thought:

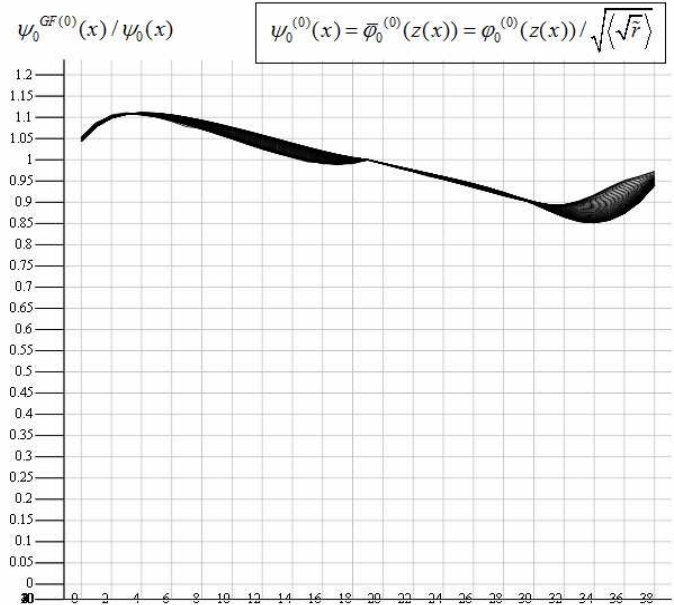

After that I tried to normalize the whole $\psi_0^{GF}(x)$ (16) as the exact $\psi_0(x)$, i.e., with the weighting function $r(x)$. I calculated the corresponding integral $N^2 = \int_0^1 \psi_0^{GF}(x)^2 r(x)dx$ numerically, not analytically, and obtained the following, more "centred" results:

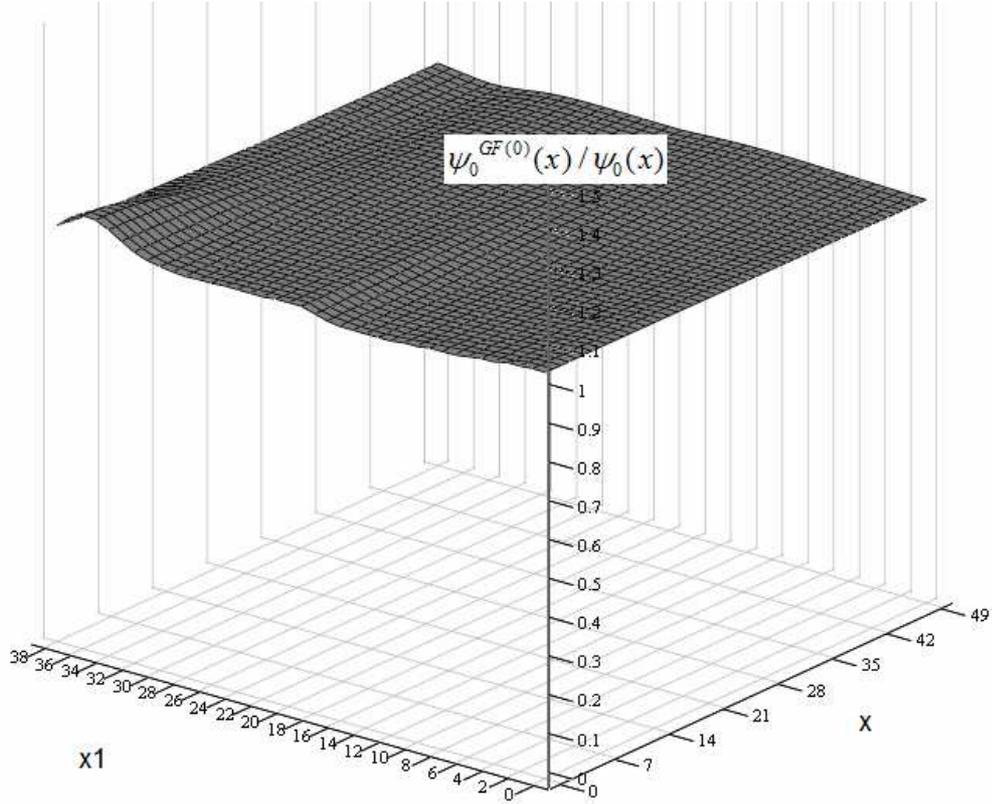

**Fig. 5.** The ratio $\psi_0^{GF(0)}(x)\,/\,\psi_0(x)$ with (16) for a normalized numerically $\psi_0^{GF(0)}(x)$.

The ratio profile picture is the following:

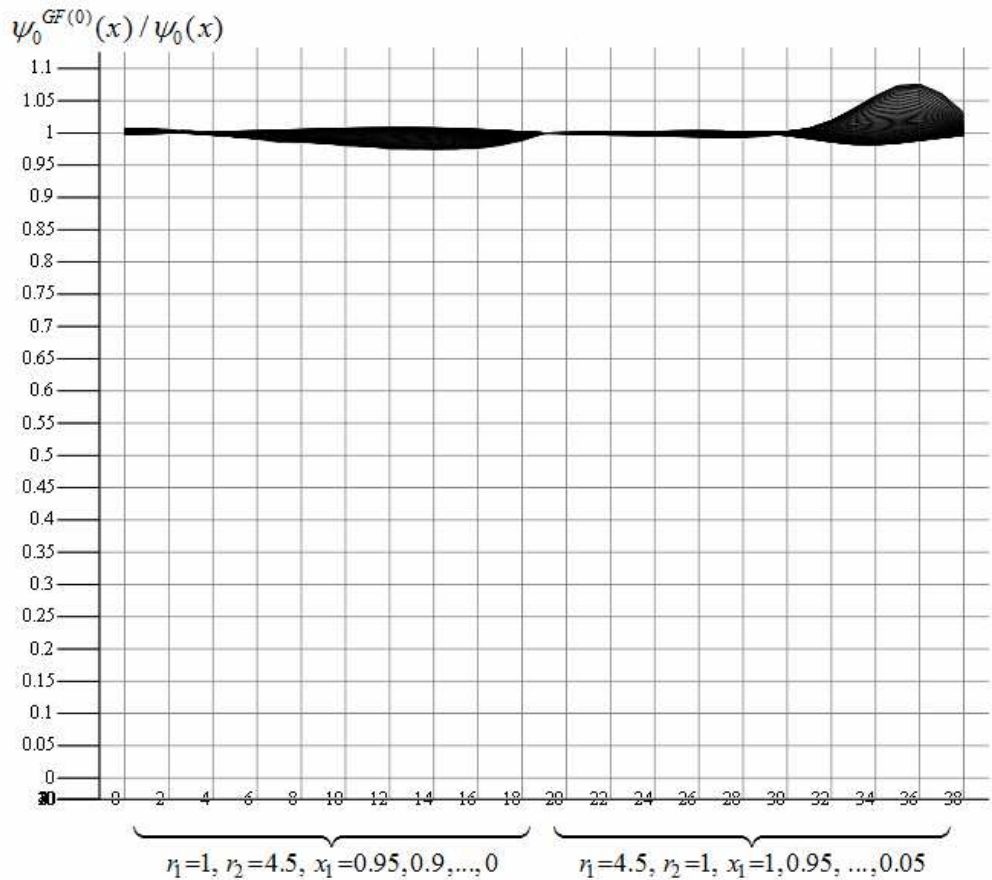

**Fig. 6.** The ratio profile $\psi_0^{GF(0)}(x)\,/\,\psi_0(x)$ with (16) for a normalized numerically $\psi_0^{GF(0)}(x)$.

Thus, formula (16) also solves the problem with practically the same precision.

Of course, the normalization integral could be calculated analytically, but it would complicate the total analytical formula, unlike the case with the WKB initial approximation, so I limited myself to a numerical calculation of normalized $\psi_0^{GF(0)}(x)$.

However, in order not to involve numerical integrations and avoid long analytical formulae, I also tried $\phi_0^{(0)}$ normalized with the original $r(x)$: $N_0^2 = \int_0^1 \phi_0^{(0)}(x)^2 r(x) dx = \sqrt{r_2} + \left(\sqrt{r_1} - \sqrt{r_2}\right)\left[\frac{z_1}{z_b} - \frac{1}{\pi}\sin\left(\pi\frac{z_1}{z_b}\right)\right]$.

This formulae is analytically shorter that that for $\psi_0^{GF(0)}(x)$. The normalization factor now depends on $\phi_0^{(0)}(x)$, unlike $\left\langle\sqrt{r}\right\rangle$. Then the Green's sum rule formula (14) gives:

$$\psi_0^{GF}(x) = \lambda_0^{FG(1)}\frac{4z_b^2}{\pi^2 N_0}\sqrt{\frac{2}{z_b}}\left[\sin\left(\frac{\pi}{2}\frac{z(x)}{z_b}\right) + \frac{\pi}{2z_b}\cos\left(\frac{\pi}{2}\frac{z_1}{z_b}\right)G_0(x, x_1)\left(\sqrt{r_2} - \sqrt{r_1}\right)\right]. \qquad (17)$$

Surprisingly, formula (17) turned out to be nearly as good as the previous formula (16):

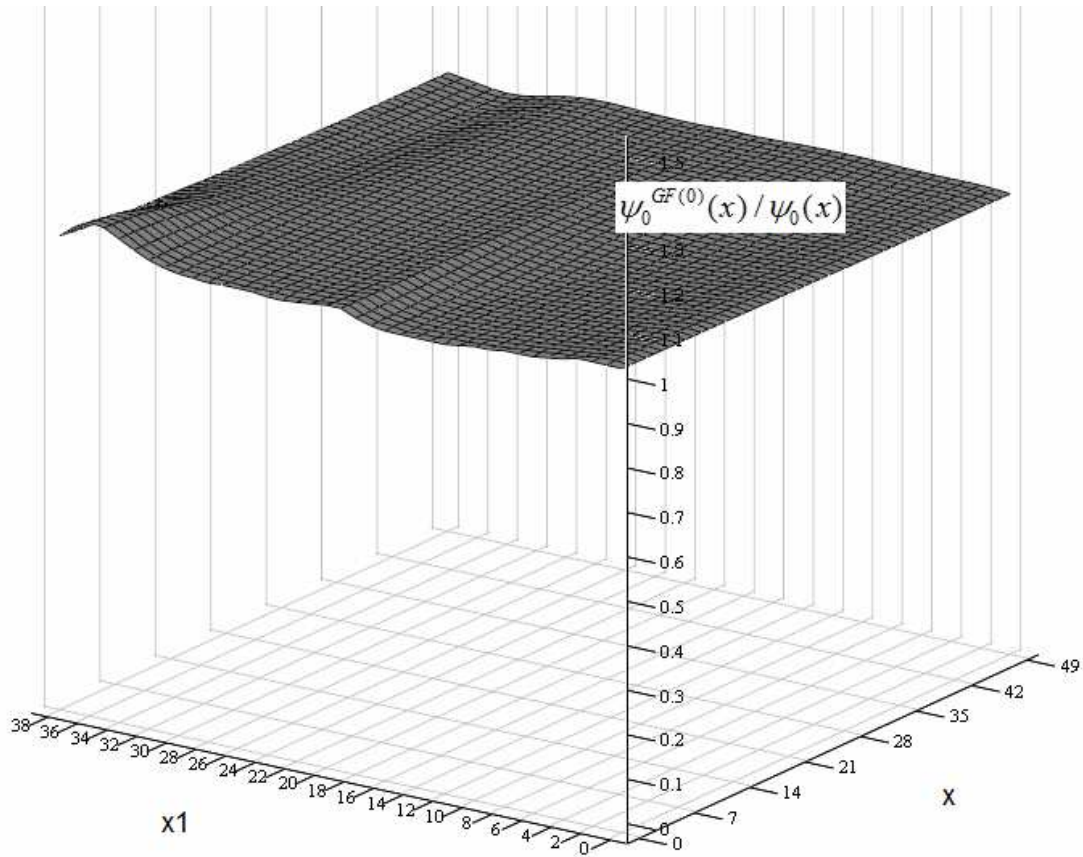

**Fig. 7.** The ratio $\psi_0^{GF(0)}(x)/\psi_0(x)$ with (17) (i.e., with normalized analytically $\phi_0^{(0)}(x)$).

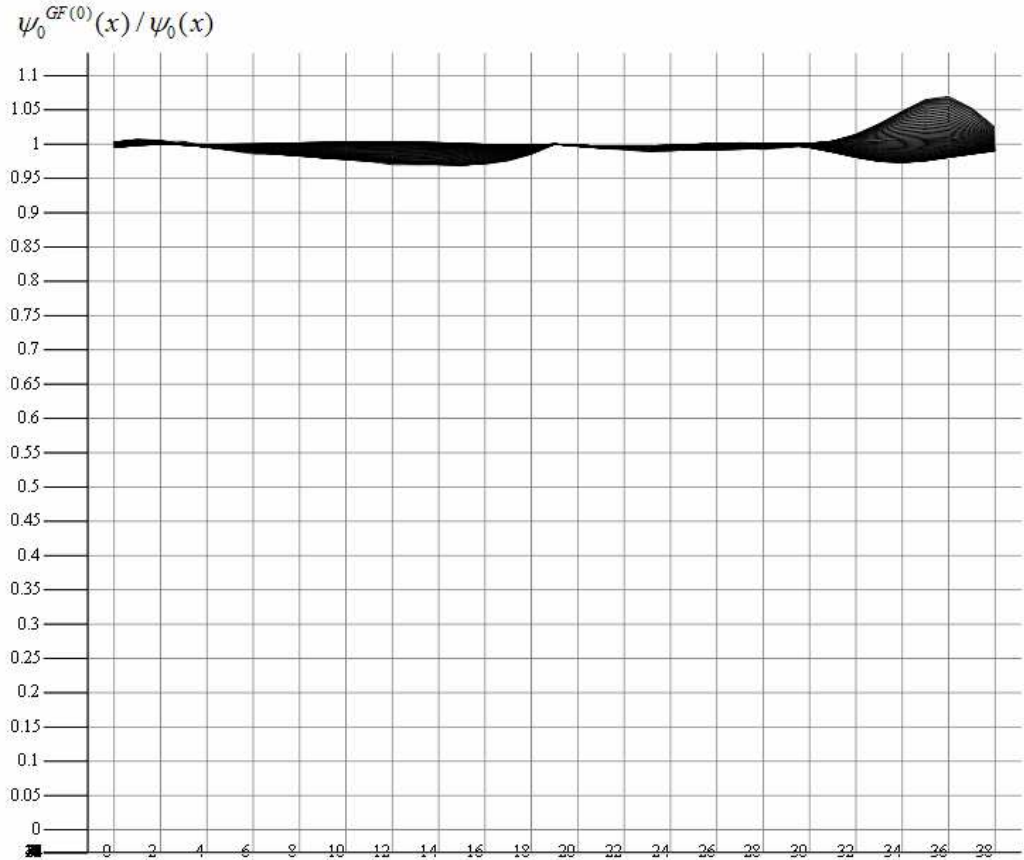

**Fig. 8.** The ratio profile $\psi_0^{GF(0)}(x) / \psi_0(x)$ with (17) (i.e., with normalized analytically $\phi_0^{(0)}(x)$ ).

Thus, formula (17) also solves the problem analytically with practically the same precision.

Above I demonstrated the precision of my analytical constructions for a rather big difference between $r_1$ and $r_2$ in the system layers - big from the point of view of perturbation theory [1]: $\varepsilon = \ln(r_1 / r_2) = \pm 1.5$, $\varepsilon / 2 = \pm 0.75$ ). For smaller differences the precision will be much better. The obtained precision turns out to be good because contribution of spectral terms with $n \geq 1$ is small not only due to effective orthohonality of $\psi_0^{(0)}$ to $\psi_{n \geq 1}$, but also due to relatively big denominators: $\propto (2n+1)^2 = 9$, 25, 49, etc. in comparison with the main term $\propto 1$ (i.e., **all suppression factors are small**: $\lambda_0 / \lambda_{n \geq 1}$ <<1: 1/9 $\approx$ 0.11, 1/25=0.04, 1/49$\approx$ 0.02, etc.).

Of course, this "equivalence" of different analytical constructions can only be established **after** comparison with the exact solution, i.e., after numerical calculations, "a posteriori". "A priori" these formulae are only good for "infinitesimal" relative difference between the neighbouring layers $\left| \ln(r_1 / r_2) \right| \ll 1$.

## 6. Next iteration

The sum rule (11) is an "implicit" construction with respect to $\psi_0(x)$. It is very important to insert a good initial approximation $\psi_0^{(0)}(x)$ into the right-hand side. (We will see it later on, while calculating $\psi_1(x)$.) Our analytical constructions (15) and (17) are even better candidates in this respect than a WKB-like approximation (2), so me may try to inject them into (14) to obtain a next approximation $\psi_0^{GF(1)}(x)$. This is an iterative procedure that may converge due to a better and better "orthonormality" of the trial $\psi_0^{(0)}$ and exact $\psi_{n \geq 1}$ eigenfunctions. Indeed, both formulae – (15) and (17) have given the following numerical results:

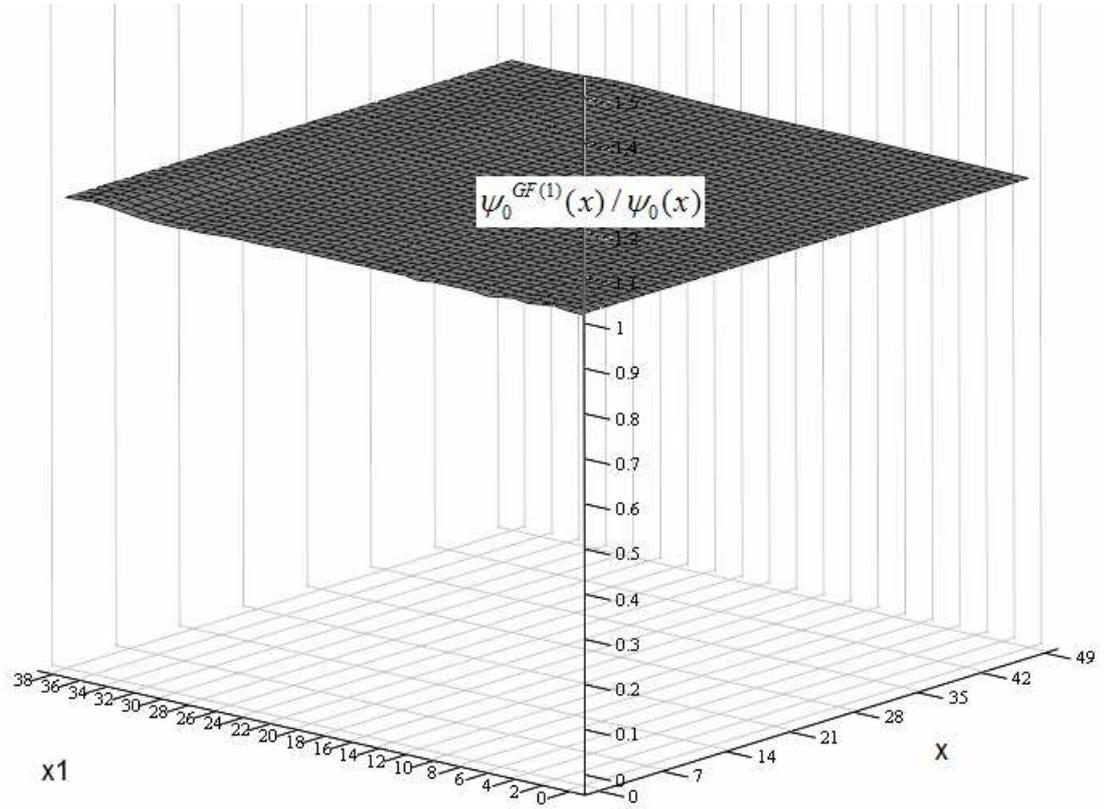

**Fig. 9.** The ratio $\psi_0^{GF(1)}(x)/\psi_0(x)$ with $\psi_0^{(0)}(x)$ in (14) from (15) (WKB).

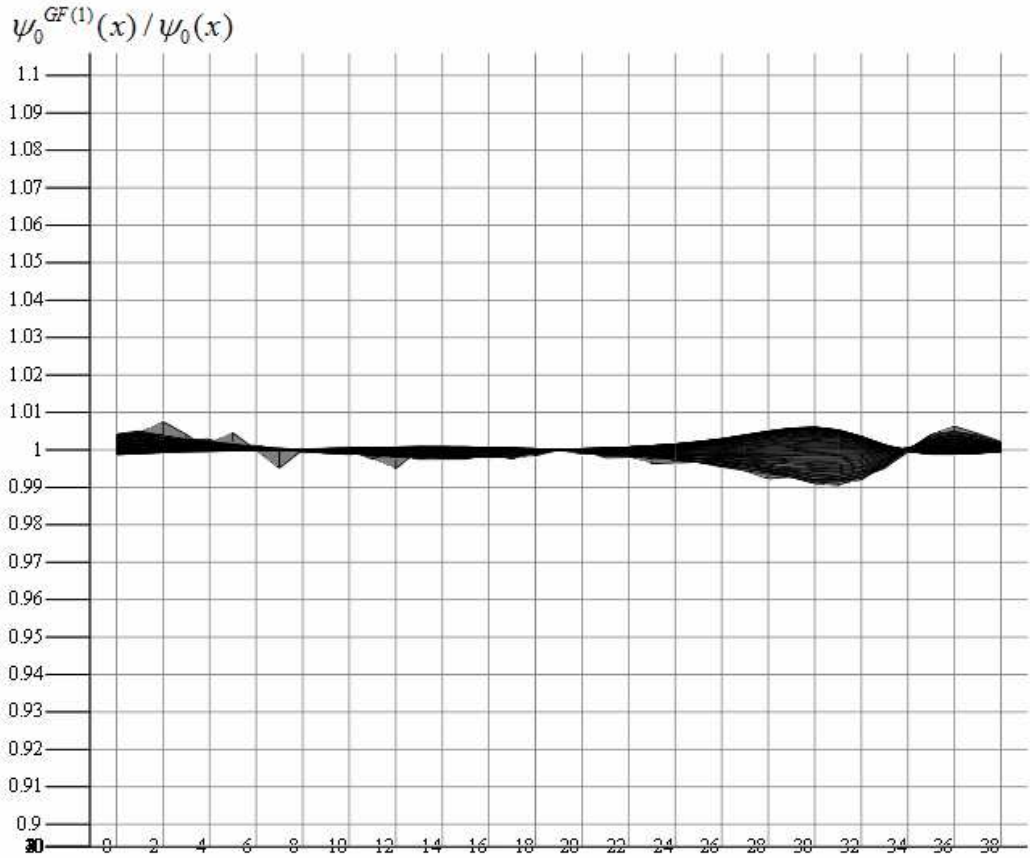

**Fig. 10.** The ratio profile of $\psi_0^{GF(1)}(x)/\psi_0(x)$ with $\psi_0^{(0)}(x)$ in (14) from (15) (WKB).

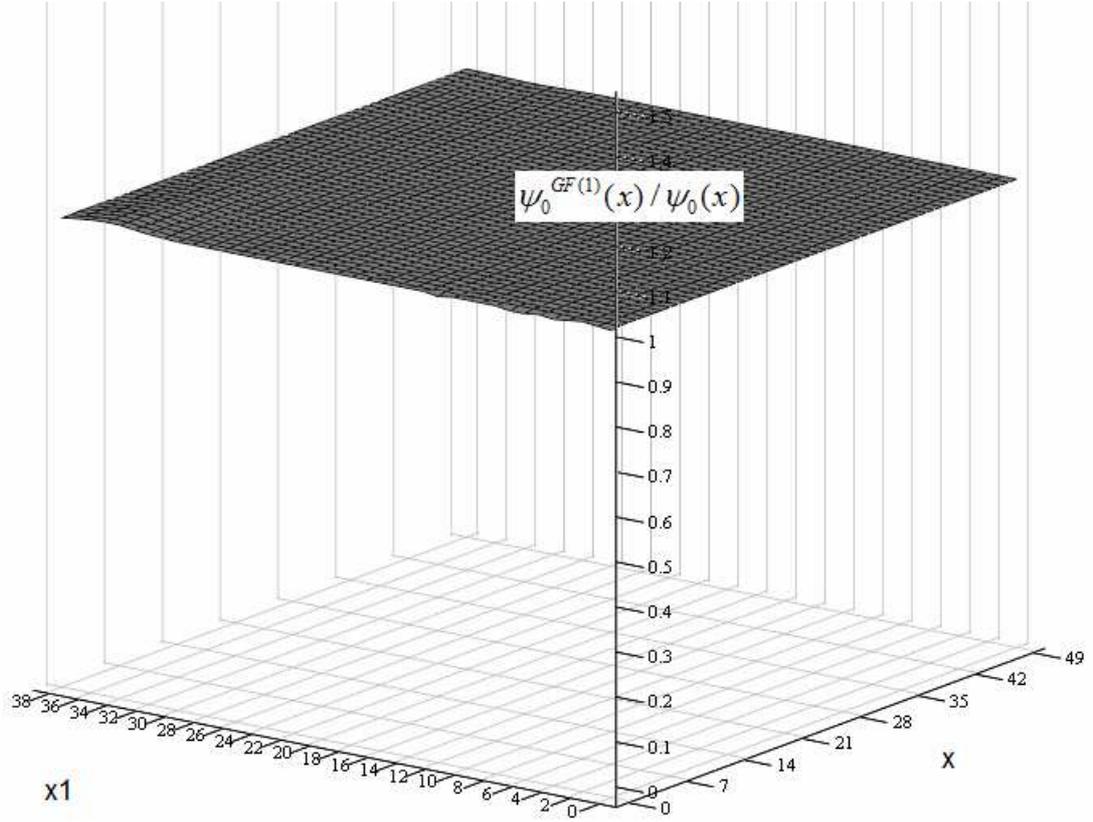

**Fig. 11.** The ratio $\psi_0^{GF(1)}(x)\,/\,\psi_0(x)$ with $\psi_0^{(0)}(x)$ in (14) from (17) (No WKB).

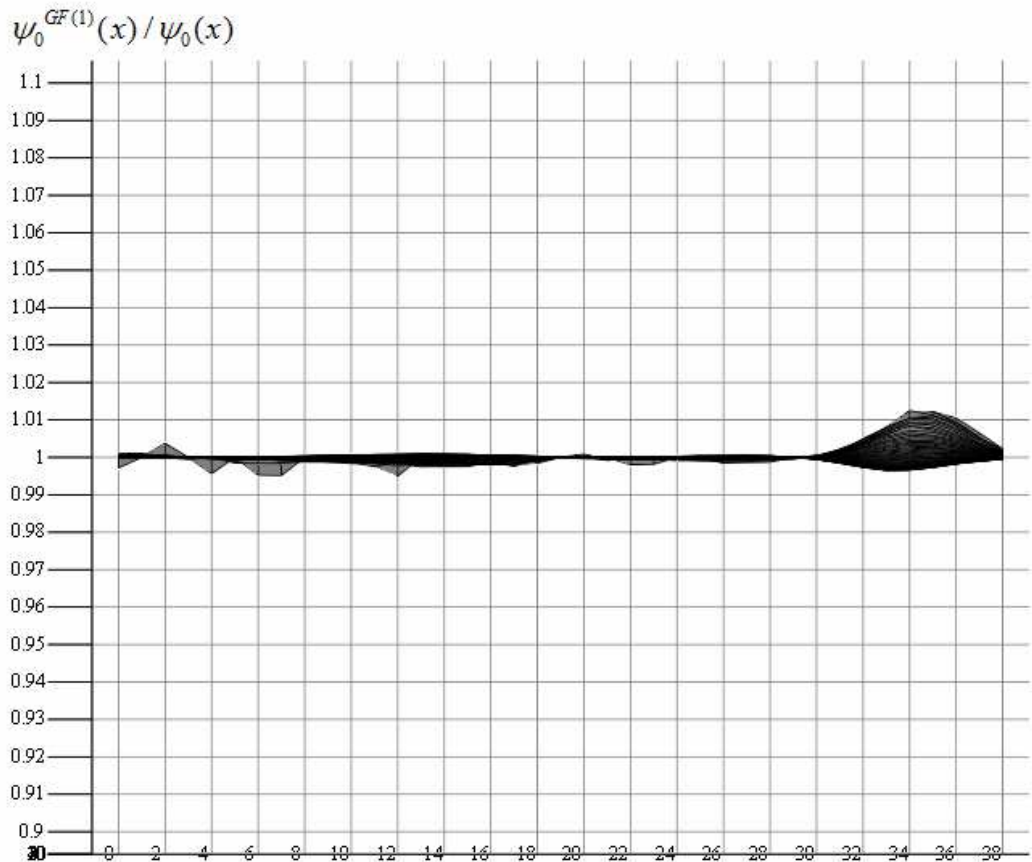

**Fig. 12.** The ratio profile of $\psi_0^{GF(1)}(x)\,/\,\psi_0(x)$ with $\psi_0^{(0)}(x)$ in (14) from (17) (No WKB).

In both cases ("WKB" (15) and "No WKB" (17) on the right-hand side of (11)) the resulting $\psi_0{}^{GF(1)}(x)$ were calculated and normalized numerically because I was not interested in putting here long analytical formulae. The relative error of $\psi_0{}^{GF(1)}(x)$ became generally much smaller and it ranges within $\pm 1\%$.

Keeping in mind that experimentally the system properties $r(x)$ (diffusion coefficient, heat conduction coefficient, density, specific heat capacity, etc.) are known with a limited precision, the solutions $\psi_0{}^{GF(1)}(x)$ might be quite sufficient for practical applications. But from a theoretical point of view this iterative procedure may converge/diverge depending on the initial approximation quality (see **Appendix 3** for a formal proof). A bright example of a "bad" initial approximation in an unstable iterative procedure is considered below.

## 7. Estimations of $\psi_1(x)$ from the Green's function $G_0$

The sum rule for $\psi_1(x)$ or a formal solution to (1) literally reads:

$$\psi_1(x) = \lambda_1 \cdot \int_0^1 r(y) \cdot \psi_1(y) \cdot G_0(x, y) dy \tag{18}$$

If the trial function $\psi_1(y)^{(0)}$ is very close to the exact one, this formula may "work" just like our previous formulae for $\psi_0(x)^{GF}$, but we start from a WKB-approximation $\psi_1{}^{(0)}(x) = \varphi_1{}^{(0)}(z(x)) / \sqrt[4]{\tilde{r}(z(x))}$ and this makes the formula too inexact. Namely, I have found that there is an essential contribution from the ground state eigenfunction scalar product $\left(\psi_1{}^{(0)}, \psi_0\right) \neq 0$ "amplified" by a relatively big coefficient $\lambda_1 / \lambda_0 \approx 10$ :

$$I_1(x) = \lambda_1 \cdot \int_0^1 r(y) \cdot \psi_1(y)^{(0)} \cdot G_0(x, y) dy = \lambda_1 \cdot \int_0^1 r(y) \cdot \psi_1(y)^{(0)} \cdot \left[ \frac{\psi_0(x)\psi_0(y)}{\lambda_0} + \sum_{n=1}^{\infty} \frac{\psi_n(x)\psi_n(y)}{\lambda_n} \right] dy =$$

$$= \underbrace{\frac{\lambda_1}{\lambda_0} \cdot \psi_0(x) \int_0^1 r(y) \, \psi_1(y)^{(0)} \psi_0(y) dy}_{\text{const} \cdot \psi_0(x)} + \underbrace{\lambda_1 \int_0^1 r(y) \cdot \psi_1(y)^{(0)} \sum_{n=1}^{\infty} \frac{\psi_n(x)\psi_n(y)}{\lambda_n} dy}_{\psi_1^{GF}(x)} . \tag{19}$$

As the contribution of $\psi_0(x)$ in (19) may be important, I defined $\psi_1(x)^{GF}$ as $I_1(x)$ with one subtraction:

$$\psi_1(x)^{GF} = I_1(x) - \psi_0(x) \frac{\lambda_1}{\lambda_0} \int_0^1 r(y)\psi_1(y)^{(0)}\psi_0(y) dy = \lambda_1 \int_0^1 r(y)\psi_1(y)^{(0)} \left[ G_0(x, y) - \frac{\psi_0(x)\psi_0(y)}{\lambda_0} \right] dy . \tag{20}$$

Odd number Figures 13-31 show the comparison of the exact eigenfunction $\psi_1(x)$ (black line) with the initial (WKB-like) approximation $\psi_1{}^{(0)}$ (green line) and functions (19) (red line) and (20) (blue line). One can see that the initial approximation is not very good since it gives an essential projection on $\psi_0(x)$ seen as the difference between the red and the black curves. Subtracting this projection from (19) renders a good approximation for $\psi_1(x)^{GF}$ (blue curves) that may be used for further iterations. Otherwise the iteration procedure (18) diverges for our big differences $\Delta r / r$ (see **Appendix 3** for a formal proof).

I tried this formula first with the exact (and a priori unknown) solutions $\psi_0(x), \lambda_0, \lambda_1$ and I found no essential difference while using the approximate (known) constructions $\psi_0{}^{GF(1)}(x)$, $\lambda_0{}^{GF(1)}$, $\lambda_1{}^{PT(1)}$ instead:

$$\psi_1{}^{GF}(x) = \lambda_1{}^{PT(1)} \cdot \int_0^1 r(y) \cdot \psi_1(y)^{(0)} \cdot G_0(x, y) dy - \psi_0(x)^{GF(1)} \cdot \frac{\lambda_1{}^{PT(1)}}{\lambda_0{}^{GF(1)}} \int_0^1 r(y) \cdot \psi_1(y)^{(0)} \cdot \psi_0(y)^{GF(1)} dy , \tag{21}$$

(compare the odd and the even number figures 13-32). So the first eigenfunction $\psi_1$ can also be estimated analytically from the Green's function sum rule (see **Appendix 2** for higher precision calculations).

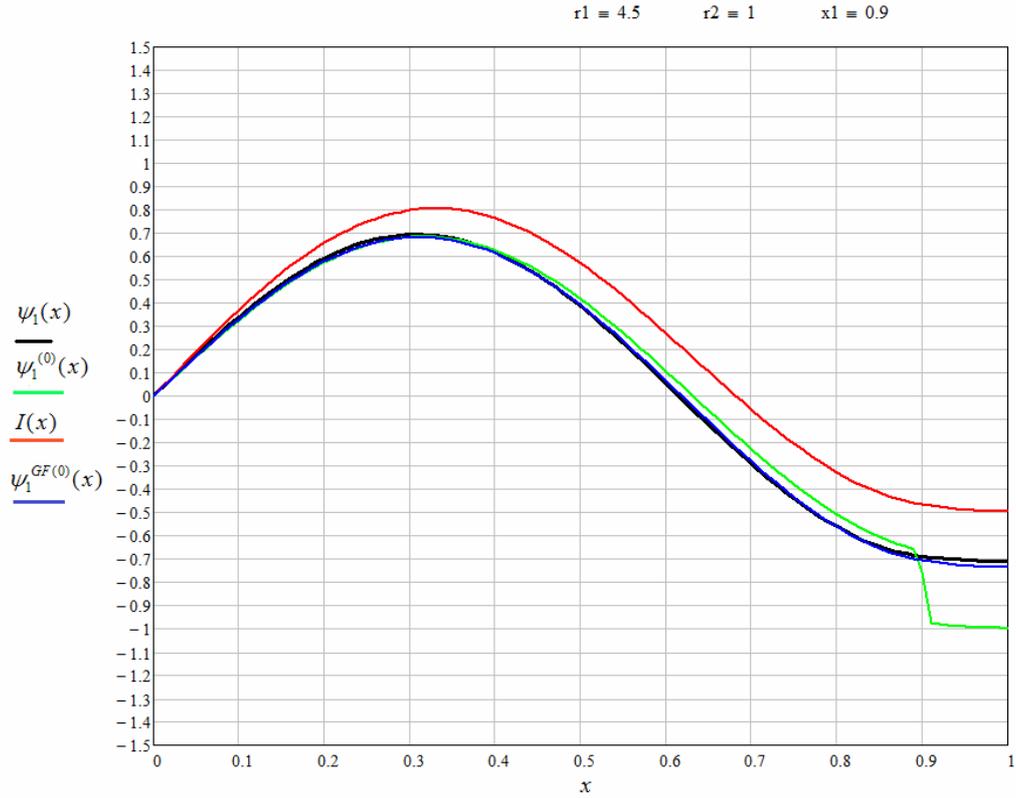

**Fig. 13.** Calculation of $\psi_1(x)^{GF(0)}$ from (20) with help of exact $\psi_0(x), \lambda_0, \lambda_1$.

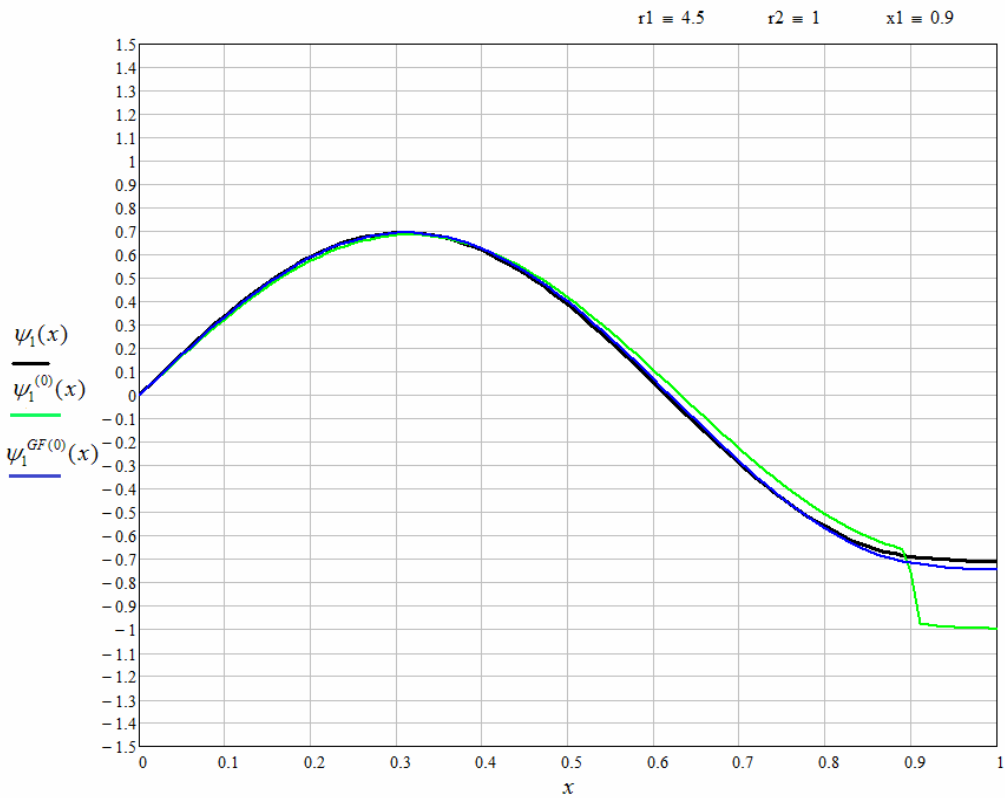

**Fig. 14.** Calculation of $\psi_1(x)^{GF(0)}$ from (21) with help of approximate $\psi_0^{GF(1)}(x), \lambda_0^{GF(1)}, \lambda_1^{PT(1)}$.

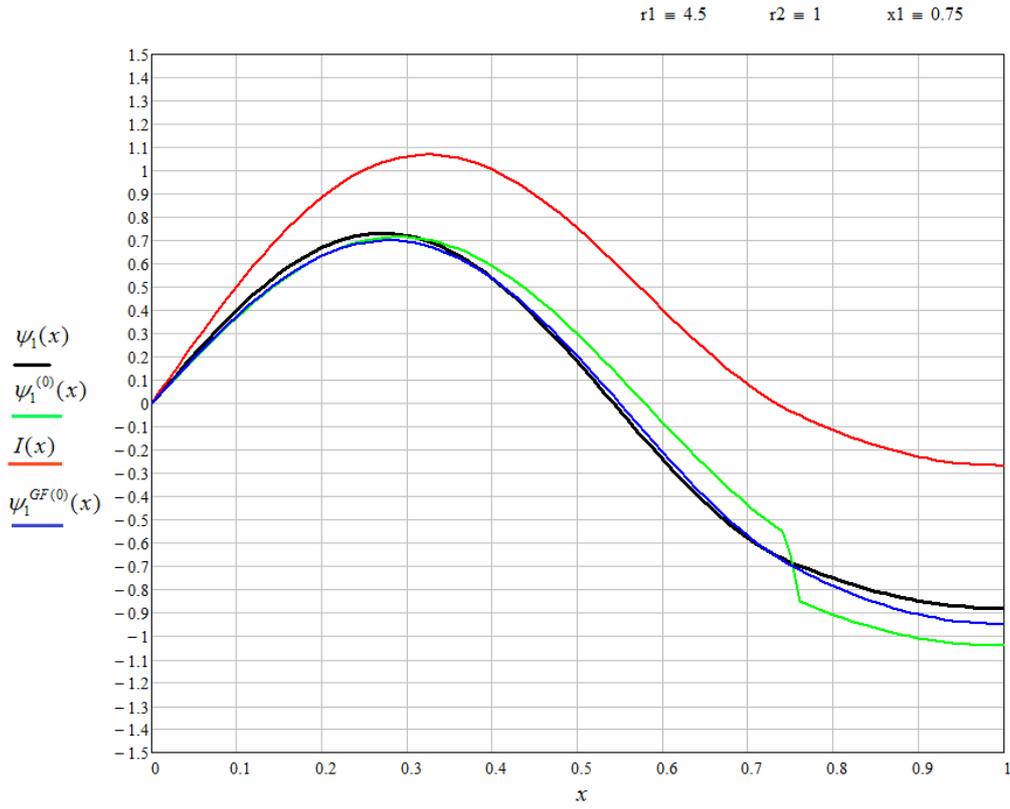

**Fig. 15.** Calculation of $\psi_1(x)^{GF(0)}$ from (20) with help of exact $\psi_0(x), \lambda_0, \lambda_1$.

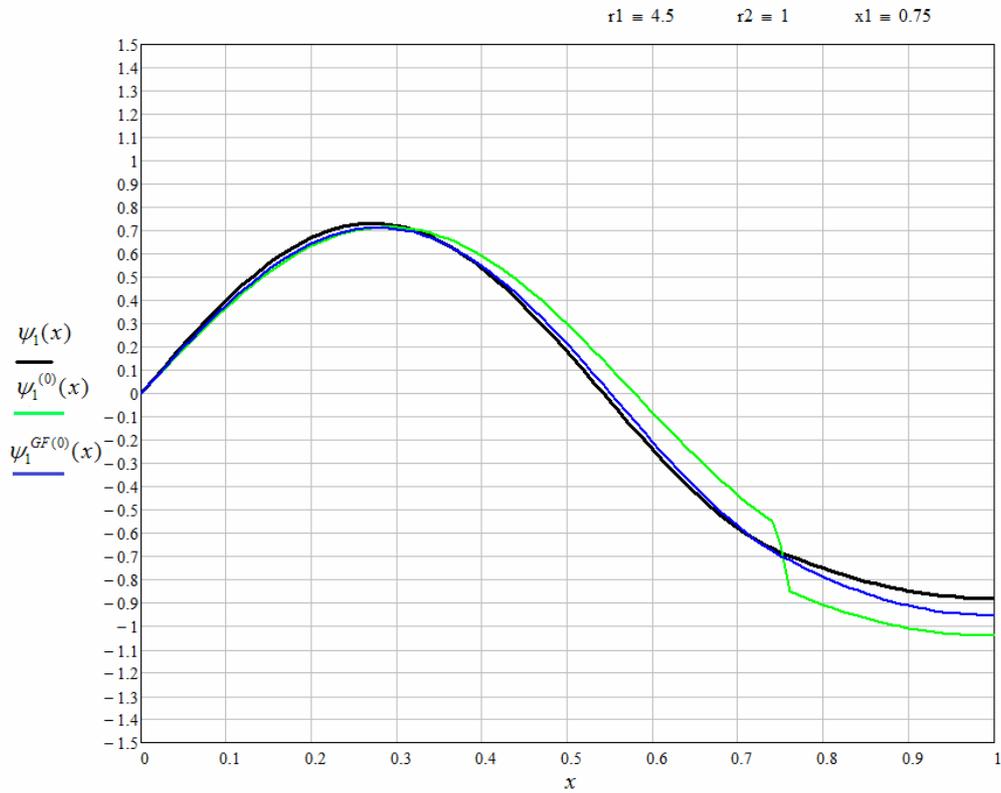

**Fig. 16.** Calculation of $\psi_1(x)^{GF(0)}$ from (21) with help of approximate $\psi_0^{GF(1)}(x), \lambda_0^{GF(1)}, \lambda_1^{PT(1)}$.

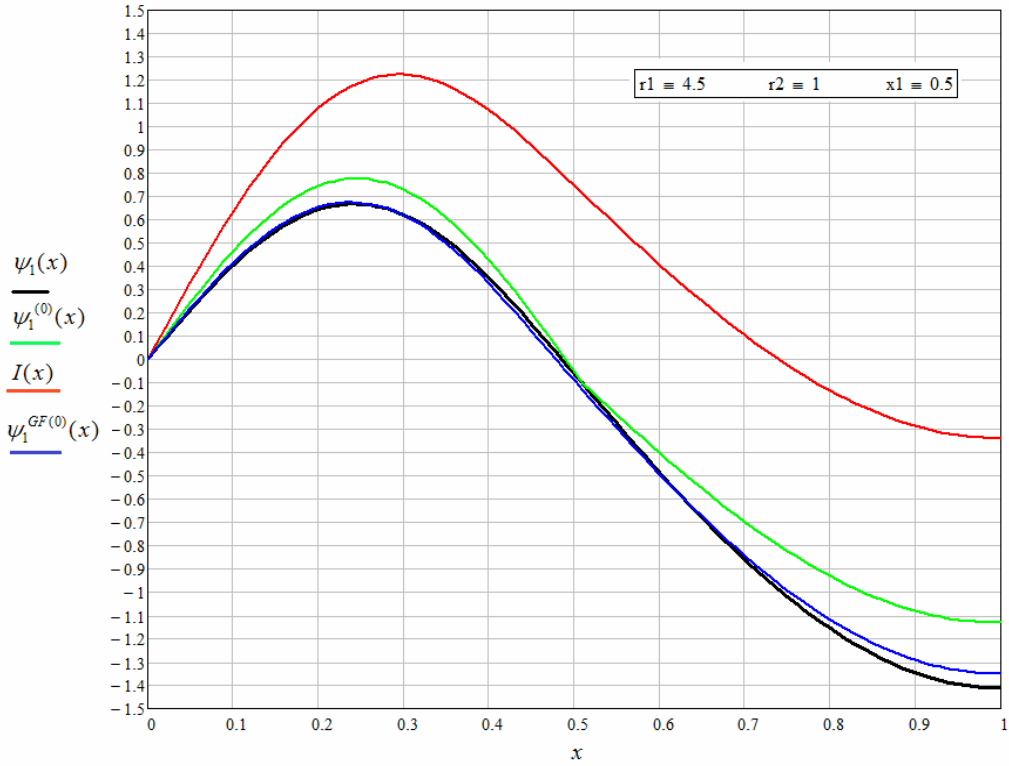

**Fig. 17.** Calculation of $\psi_1(x)^{GF(0)}$ from (20) with help of exact $\psi_0(x), \lambda_0, \lambda_1$.

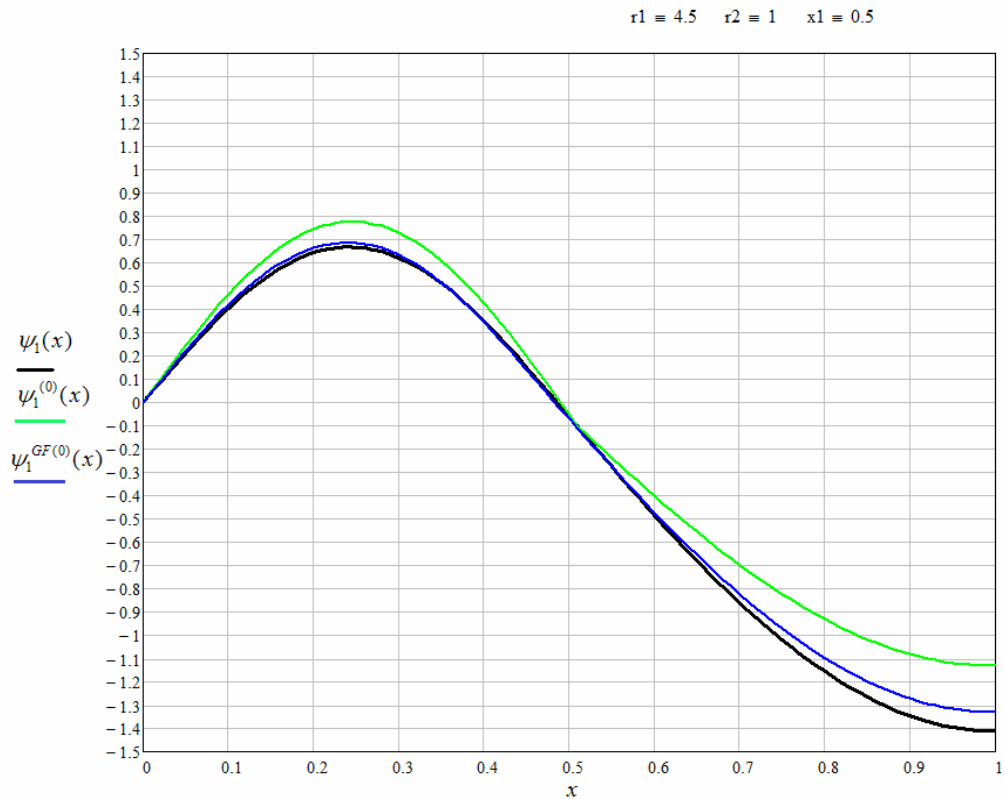

**Fig. 18.** Calculation of $\psi_1(x)^{GF(0)}$ from (21) with help of approximate $\psi_0^{GF(1)}(x), \lambda_0^{GF(1)}, \lambda_1^{PT(1)}$.

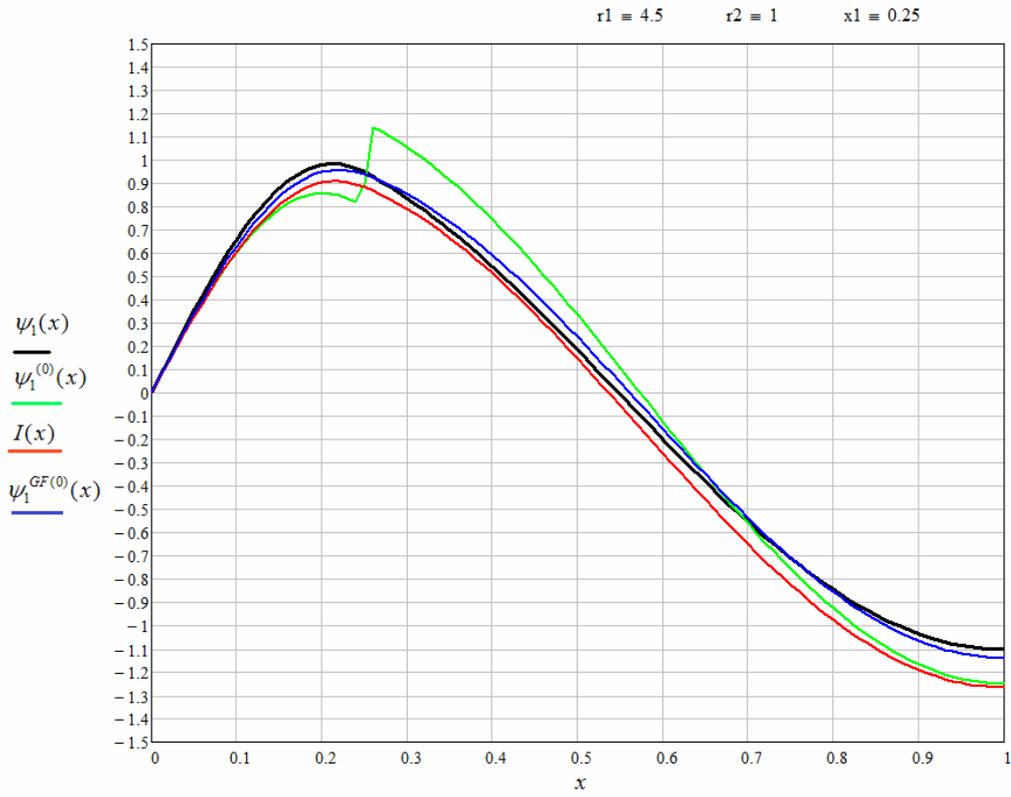

**Fig. 19.** Calculation of $\psi_1(x)^{GF(0)}$ from (20) with help of exact $\psi_0(x), \lambda_0, \lambda_1$.

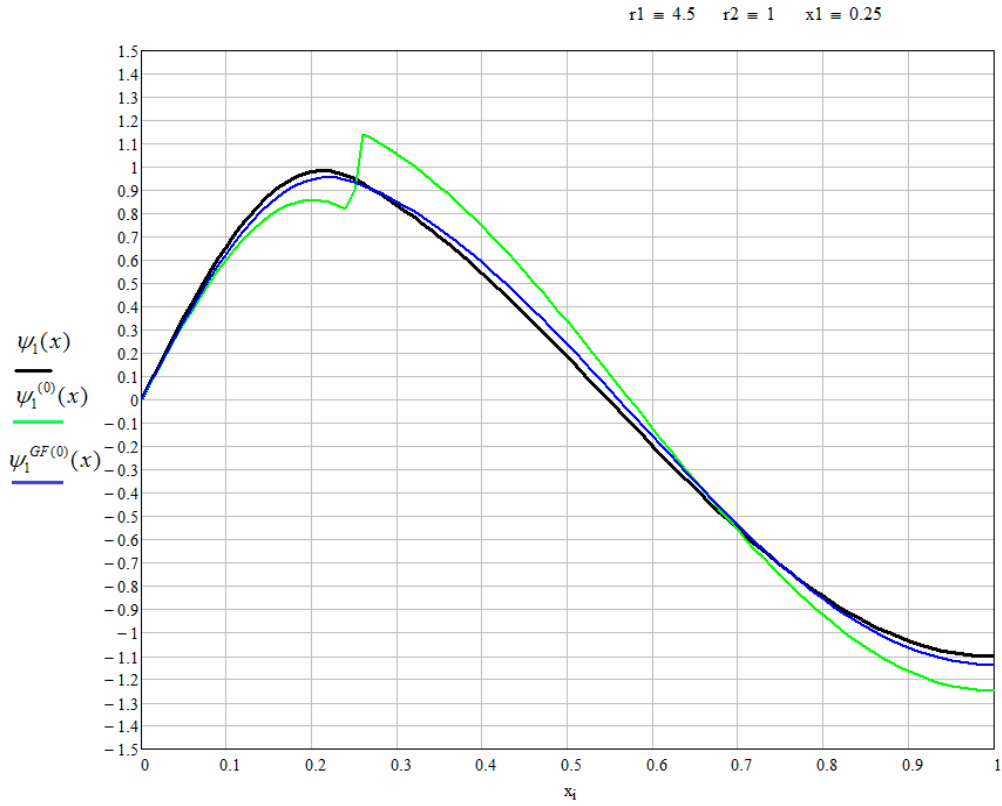

**Fig. 20.** Calculation of $\psi_1(x)^{GF(0)}$ from (21) with help of approximate $\psi_0^{GF(1)}(x), \lambda_0^{GF(1)}, \lambda_1^{PT(1)}$.

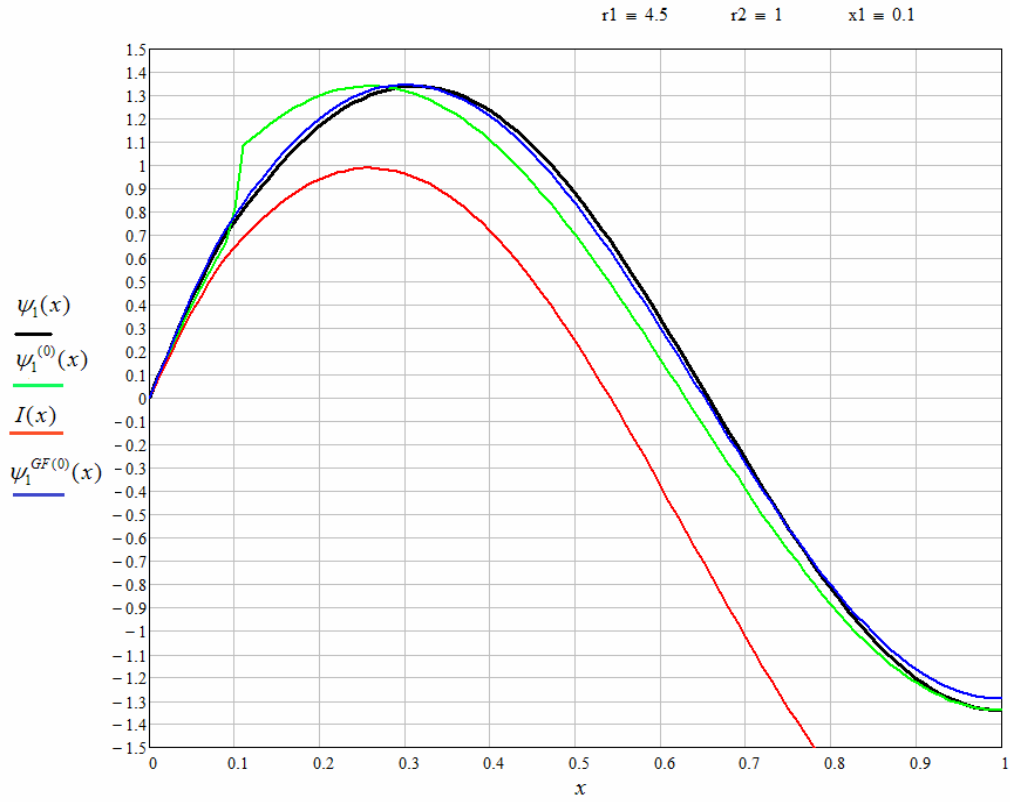

**Fig. 21.** Calculation of $\psi_1(x)^{GF(0)}$ from (20) with help of exact $\psi_0(x), \lambda_0, \lambda_1$.

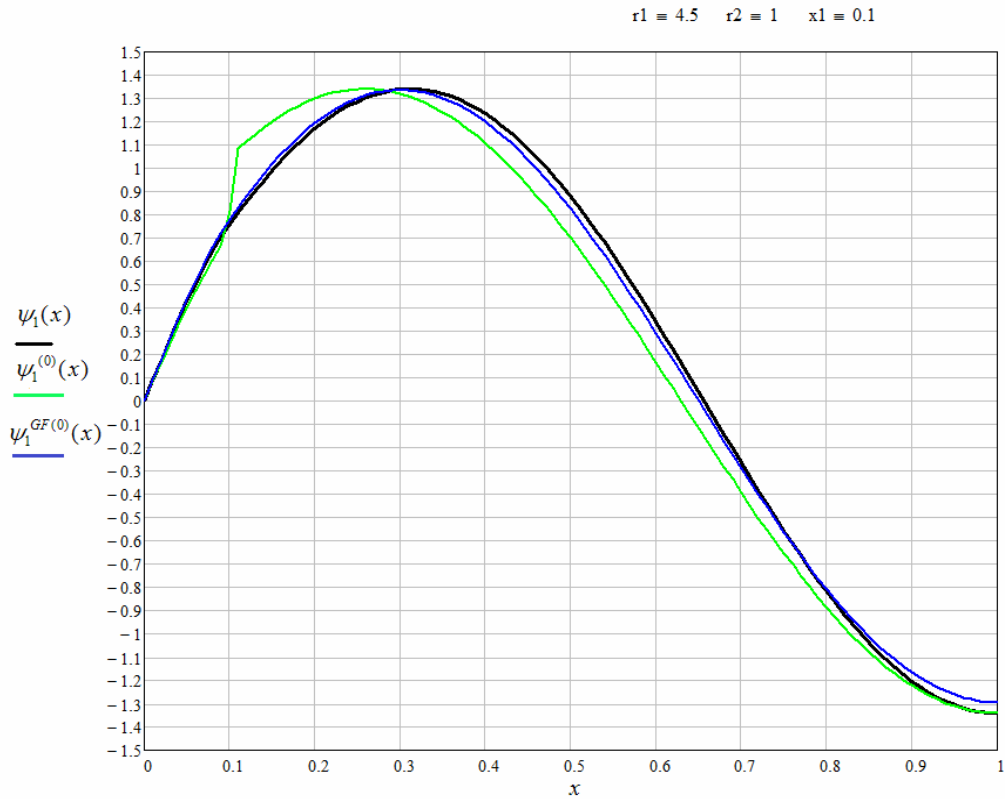

**Fig. 22.** Calculation of $\psi_1(x)^{GF(0)}$ from (21) with help of approximate $\psi_0^{GF(1)}(x), \lambda_0^{GF(1)}, \lambda_1^{PT(1)}$.

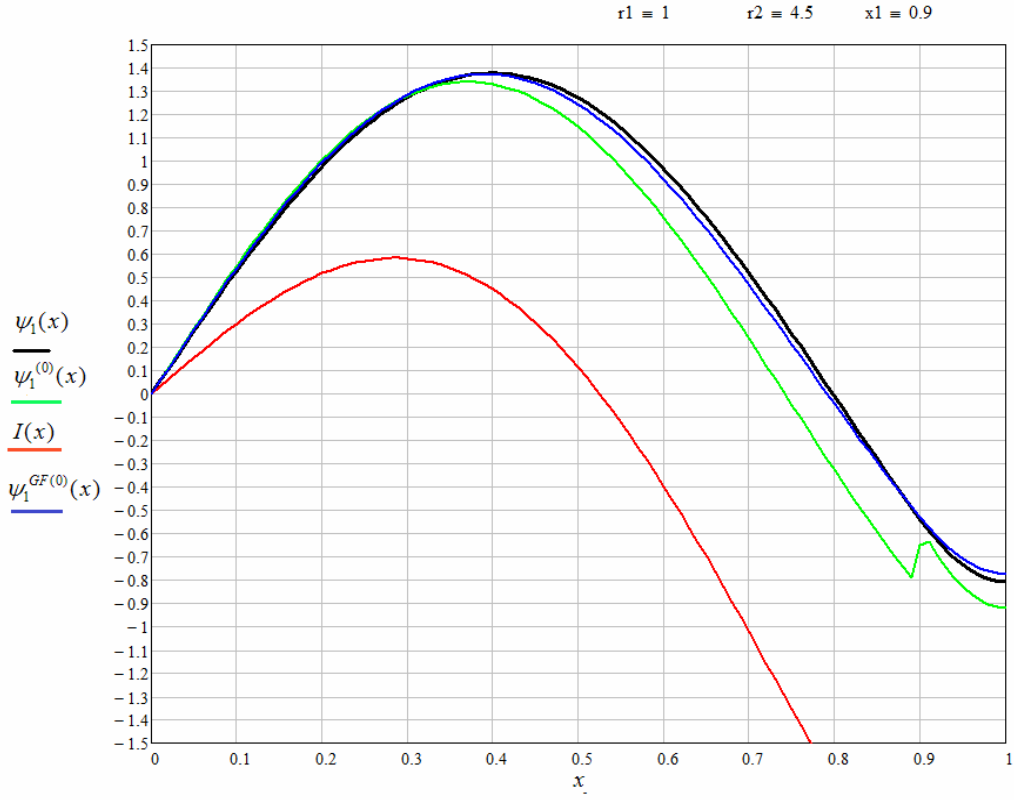

**Fig. 23.** Calculation of $\psi_1(x)^{GF(0)}$ from (20) with help of exact $\psi_0(x), \lambda_0, \lambda_1$.

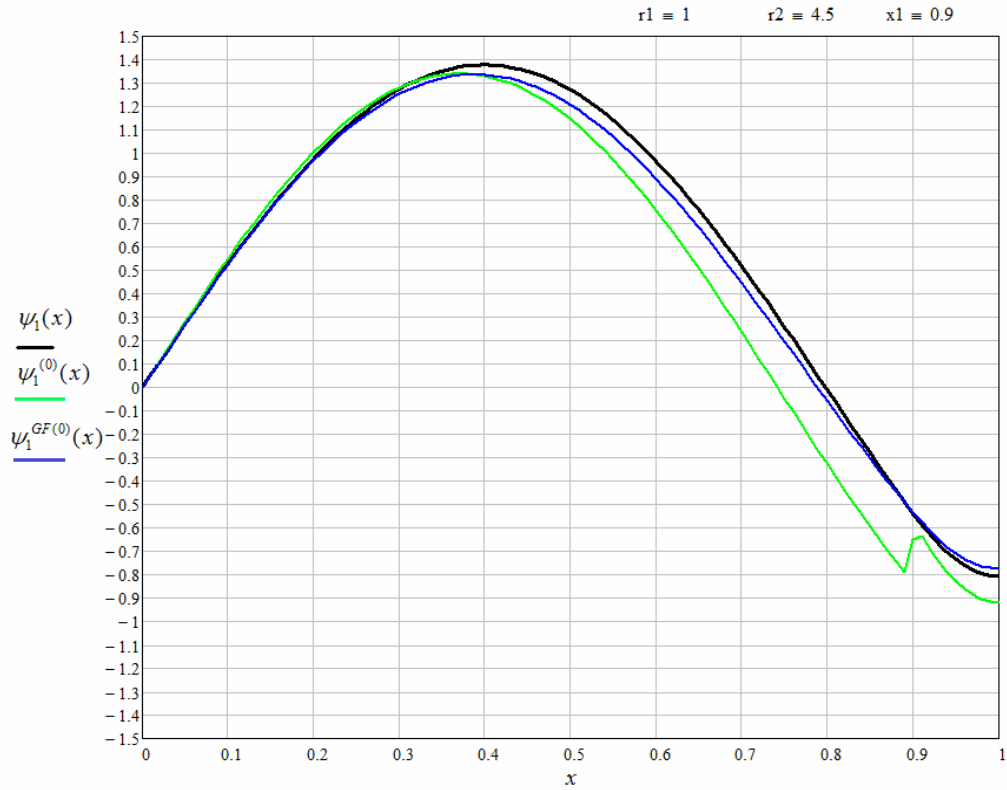

**Fig. 24.** Calculation of $\psi_1(x)^{GF(0)}$ from (21) with help of approximate $\psi_0^{GF(1)}(x), \lambda_0^{GF(1)}, \lambda_1^{PT(1)}$.

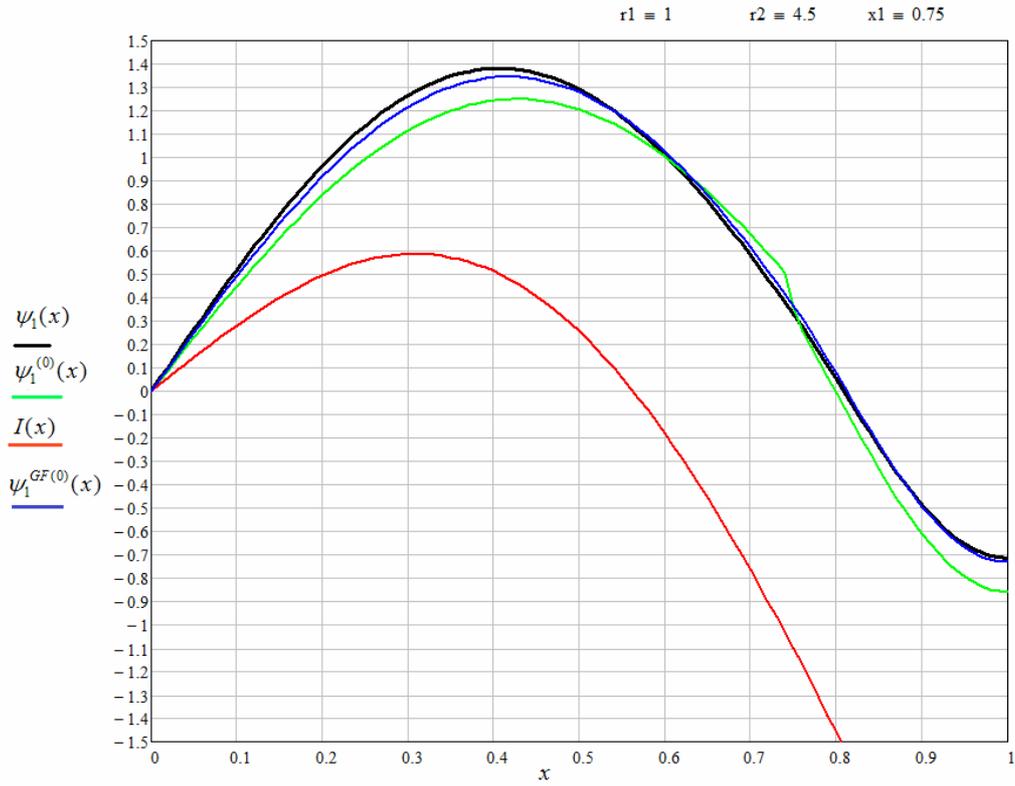

**Fig. 25.** Calculation of $\psi_1(x)^{GF(0)}$ from (20) with help of exact $\psi_0(x), \lambda_0, \lambda_1$.

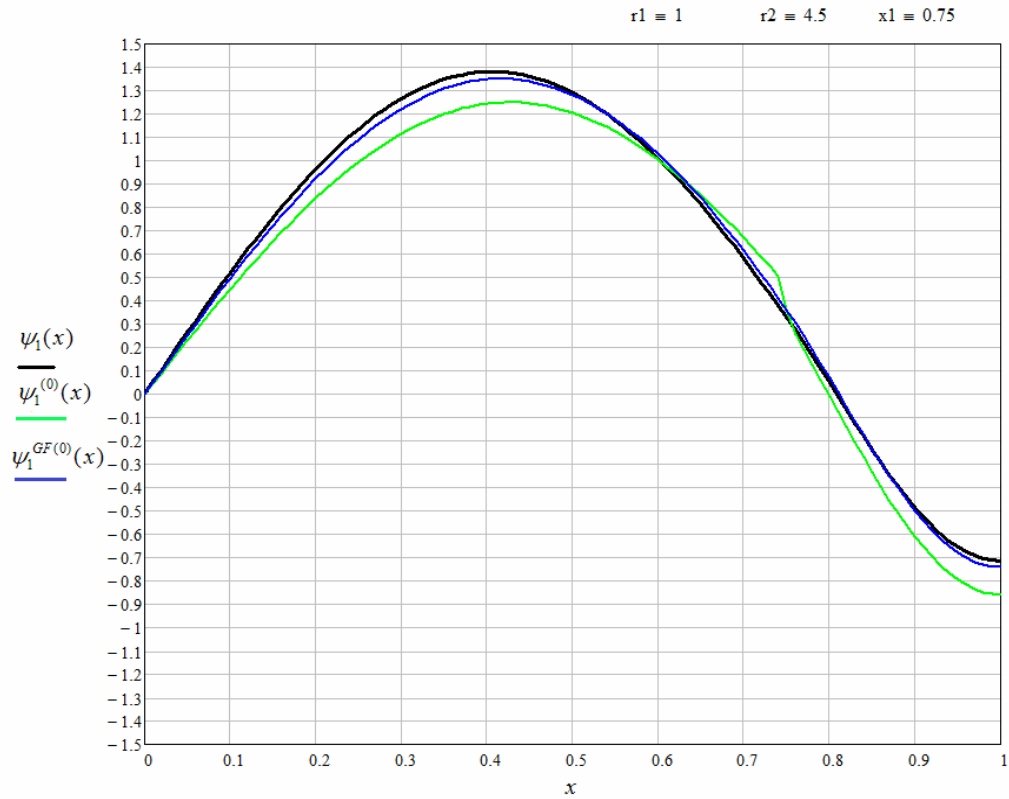

**Fig. 26.** Calculation of $\psi_1(x)^{GF(0)}$ from (21) with help of approximate $\psi_0^{GF(1)}(x), \lambda_0^{GF(1)}, \lambda_1^{PT(1)}$.

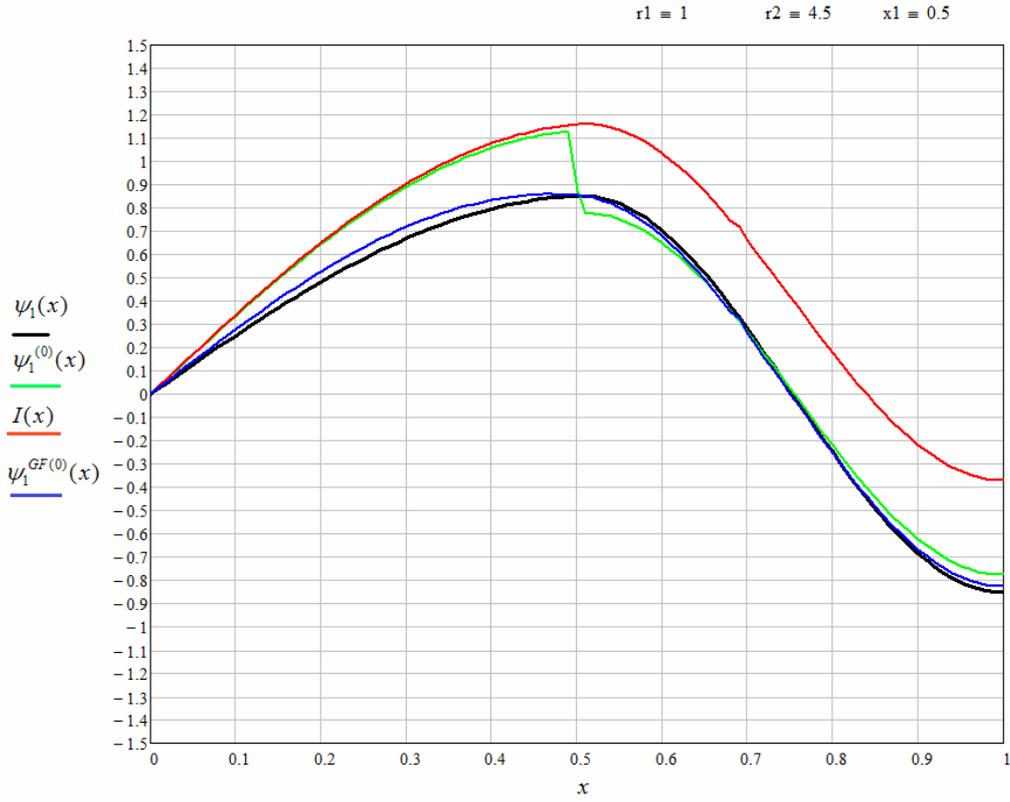

**Fig. 27.** Calculation of $\psi_1(x)^{GF(0)}$ from (20) with help of exact $\psi_0(x), \lambda_0, \lambda_1$.

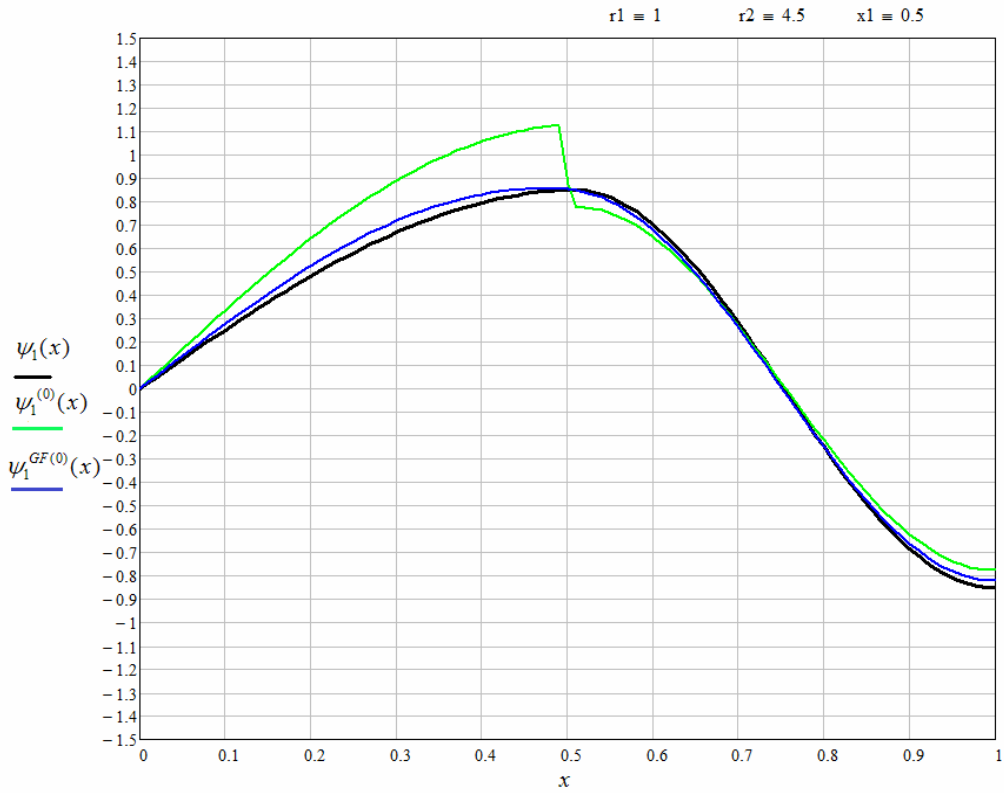

**Fig. 28.** Calculation of $\psi_1(x)^{GF(0)}$ from (21) with help of approximate $\psi_0^{GF(1)}(x), \lambda_0^{GF(1)}, \lambda_1^{PT(1)}$.

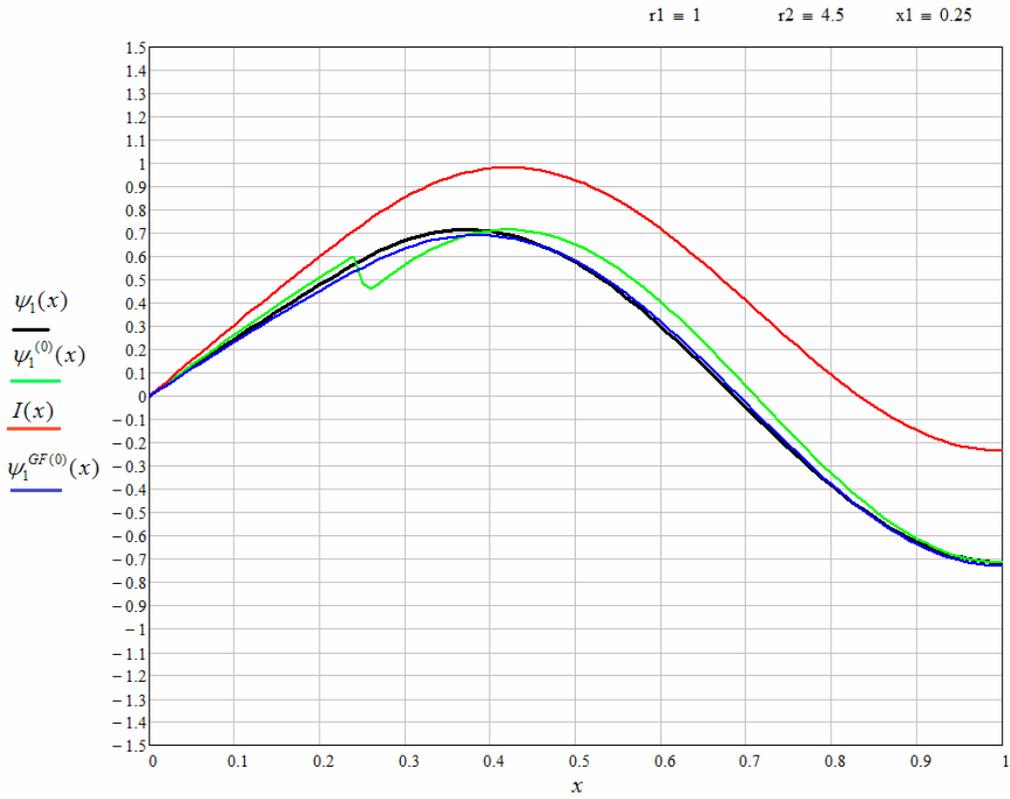

**Fig. 29.** Calculation of $\psi_1(x)^{GF(0)}$ from (20) with help of exact $\psi_0(x), \lambda_0, \lambda_1$.

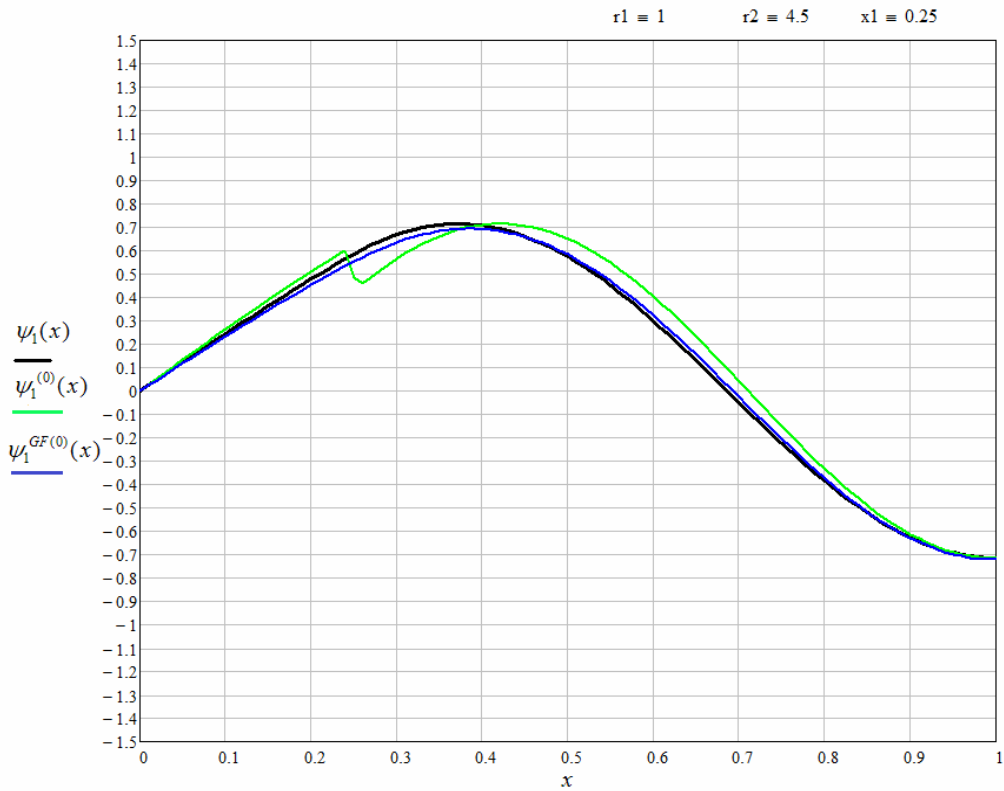

**Fig. 30.** Calculation of $\psi_1(x)^{GF(0)}$ from (21) with help of approximate $\psi_0^{GF(1)}(x), \lambda_0^{GF(1)}, \lambda_1^{PT(1)}$.

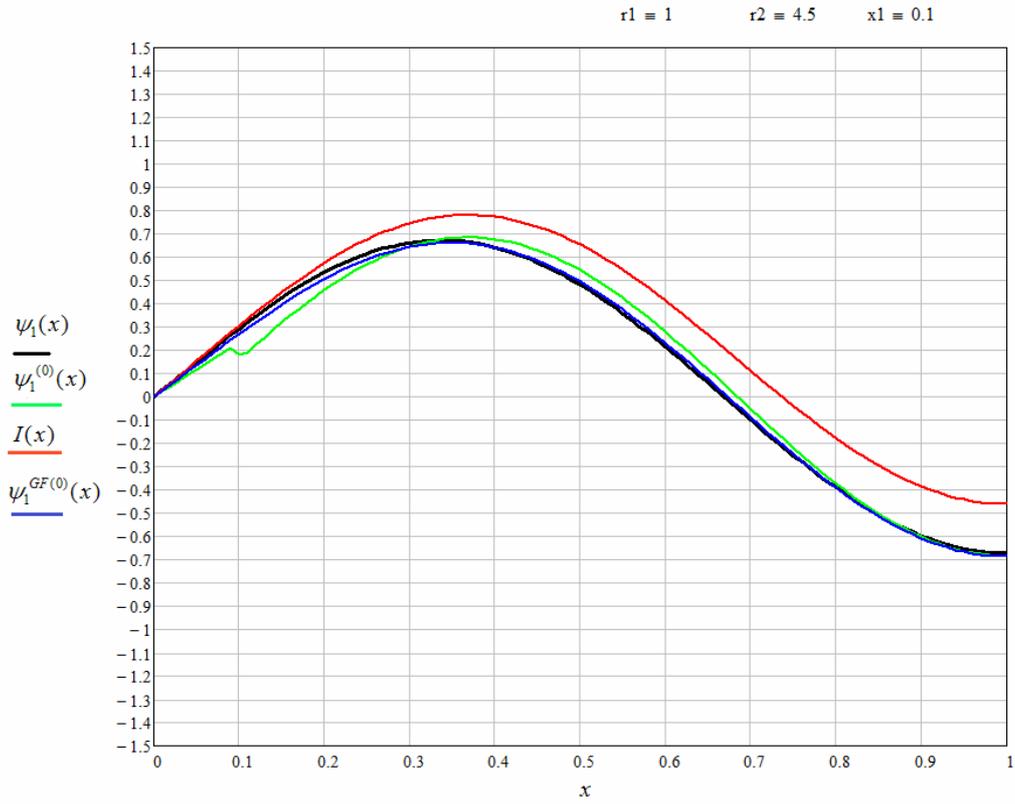

**Fig. 31.** Calculation of $\psi_1(x)^{GF(0)}$ from (20) with help of exact $\psi_0(x), \lambda_0, \lambda_1$.

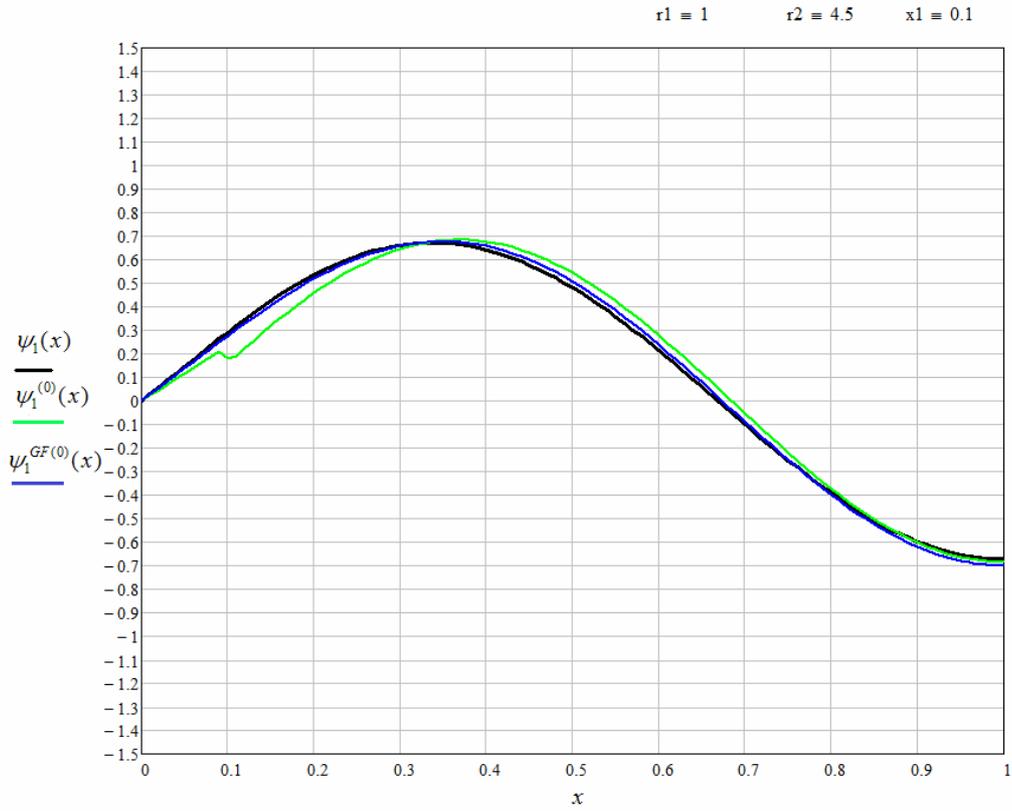

**Fig. 32.** Calculation of $\psi_1(x)^{GF(0)}$ from (21) with help of approximate $\psi_0^{GF(1)}(x), \lambda_0^{GF(1)}, \lambda_1^{PT(1)}$.

I did not plot the ratio pictures since the ratio of two functions crossing zero at slightly different points has a singularity due to zero in the denominator (see Fig. 33). The relative error may be defined differently; for example, "far from nodes", but I left the comparison above being done graphically. (Next iteration is shortly described in **Appendix 2**).

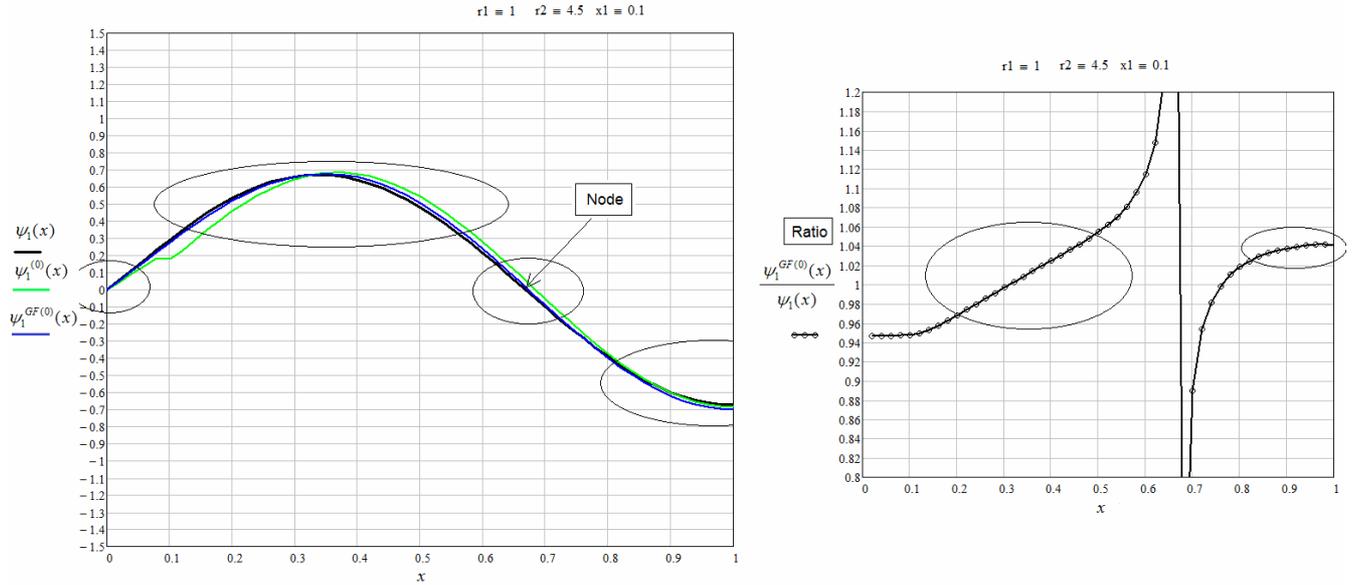

**Fig. 33.** Comparison of exact and approximate $\psi_1$. Regions around nodes $\psi_n(x) = 0$, $0 < x < 1$ are not good for making ratios (see Fig. 1 too). **The accuracy of $\psi_1^{GF}$ (21) may be somewhat worse than that of $\psi_0^{GF}$ (14) due to relatively smaller denominators with respect to the main term $\propto 9^{-1}$: for $n \geq 2$ $\lambda_1 / \lambda_{n \geq 2} \approx 9/25 = 0.36$, $9/49 \approx 0.184$, etc.**

## 8. Tests of the Perturbation Theory: the first-order eigenfunctions $\psi_n^{PT(1)}(x)$

The regular PT (**Section 2**) provides the eigenfunctions as spectral sums, which can be calculated either literally as sums over $m$ or, which is much more interesting, as shorter analytical formulae. Namely, the sum arising in the first order calculation of $\psi_n^{PT(1)}(x)$: $\text{Sum}\left(z, \lambda_n^{(0)}\right) = \sum_{m \neq n}^{\infty} \dfrac{\varphi_m^{(0)}(z)\varphi_m^{(0)}(z_1)}{\lambda_n^{(0)} - \lambda_m^{(0)}}$ can be represented as a short analytical formula due to the following identities (see (4.2) proposed in [2]):

$$\sum_{m \neq n}^{\infty} \text{Sum}\left(z, \lambda_n^{(0)}\right) = \left\{ \frac{\partial}{\partial \lambda}\left[\left(\lambda_n^{(0)} - \lambda\right)G_\lambda^{(0)}(z, z_1)\right]\right\}_{\lambda \to \lambda_n^{(0)}} = \lim_{\lambda \to \lambda_n^{(0)}}\left\{ \frac{\varphi_m^{(0)}(z)\varphi_m^{(0)}(z_1)}{\lambda_n^{(0)} - \lambda} - G_\lambda^{(0)}(z, z_1)\right\}. \quad (22)$$

Here I use an exactly constructible zeroth-order Green's function $G_\lambda^{(0)}(z, z_1)$ given below. One has to expand the $\lambda$-dependent pole functions to the second order in powers of $\varepsilon_\lambda = \lambda - \lambda_n^{(0)}$ in order to obtain a correct finite analytical expression. For our particular case the sum formula is derived as the following:

$$\frac{d^2}{dz^2}G_\lambda^{(0)} + \lambda \cdot G_\lambda^{(0)} = -\delta(z - z_1); \quad G_\lambda^{(0)}(z, z_1) = \sum_{m=0}^{\infty} \frac{\varphi_m^{(0)}(z)\varphi_m^{(0)}(z_1)}{\lambda_m^{(0)} - \lambda} = -\sum_{m=0}^{\infty} \frac{\varphi_m^{(0)}(z)\varphi_m^{(0)}(z_1)}{\lambda - \lambda_m^{(0)}} = \quad (23a)$$

$$= \frac{1}{\sqrt{\lambda}\cos\sqrt{\lambda}z_b}\begin{cases} \sin\left(\sqrt{\lambda}z\right)\cos\left(\sqrt{\lambda}(z_b - z_1)\right), \ z \leq z_1 \ , \\ \sin\left(\sqrt{\lambda}z_1\right)\cos\left(\sqrt{\lambda}(z_b - z)\right), \ z \geq z_1 \ . \end{cases} \quad (23b)$$

Here one can integrate the differential equation directly or use a general expression via linearly independent solutions and the Wronskian [3] for constructing (23b). From (22) we obtain:

$$\text{Sum}\left(z(x), \lambda_n^{(0)}\right) =$$

$$= -\frac{\sin\left(\sqrt{\lambda}z\right)\sin\left(\sqrt{\lambda}z_1\right)}{2\lambda z_b} + \frac{1}{\sqrt{\lambda}}\begin{cases} \dfrac{z}{z_b}\cos\left(\sqrt{\lambda}z\right)\sin\left(\sqrt{\lambda}z_1\right) - (1-\dfrac{z_1}{z_b})\sin\left(\sqrt{\lambda}z\right)\cos\left(\sqrt{\lambda}z_1\right),\ z \le z_1, \\[2mm] \dfrac{z_1}{z_b}\cos\left(\sqrt{\lambda}z_1\right)\sin\left(\sqrt{\lambda}z\right) - (1-\dfrac{z}{z_b})\sin\left(\sqrt{\lambda}z_1\right)\cos\left(\sqrt{\lambda}z\right),\ z \ge z_1, \end{cases},\ \lambda = \lambda_n^{(0)}.$$

<div align="right">(24a)</div>

In [2] I proposed a somewhat different, but equivalent, analytical expression, which (after correcting obvious typos in it) is the following (here $z_1 = z(x_1)$, $z_b = z(1)$ for short):

$$\mathrm{Sum}\left(z,\lambda_n^{(0)}\right) = \frac{-1}{2\sqrt{\lambda}z_b}\left\{\frac{1}{\sqrt{\lambda}}\sin\left(\sqrt{\lambda}z\right)\sin\left(\sqrt{\lambda}z_1\right) - 2\left[z_1\sin\left(\sqrt{\lambda}z\right)\cos\left(\sqrt{\lambda}z_1\right) + z\sin\left(\sqrt{\lambda}z_1\right)\cos\left(\sqrt{\lambda}z\right)\right] + \right.$$
$$\left. + z_b\left[\sin\left(\sqrt{\lambda}(z_1+z)\right) - \sin\left(\sqrt{\lambda}\left|z_1-z\right|\right)\right]\right\}_{\lambda=\lambda_n^{(0)}},$$

<div align="right">(24b)</div>

$$\psi_n^{PT(1)}(x) = \varphi_n^{PT(1)}(z(x)) = \varphi_n^{(0)}(z(x)) + \frac{1}{2}\ln\left(\frac{r_1}{r_2}\right)\cdot\varphi_n^{(0)\prime}(z(x_1))\cdot\sum_{m\ne n}\frac{\varphi_m^{(0)}(z(x))\varphi_m^{(0)}(z(x_1))}{\lambda_n^{(0)} - \lambda_m^{(0)}} =$$
$$= \sqrt{\frac{2}{z_b}}\sin\left(\sqrt{\lambda_n^{(0)}}z(x)\right) + \frac{1}{2}\ln\left(\frac{r_1}{r_2}\right)\sqrt{\frac{2\lambda_n^{(0)}}{z_b}}\cos\left(\sqrt{\lambda_n^{(0)}}z_1\right)\cdot\mathrm{Sum}\left(z(x),\lambda_n^{(0)}\right).$$

<div align="right">(25)</div>

Formula (25) generally reproduces well the exact $\psi_n(x)$ up to a normalization factor. I normalized the whole $\psi_n^{(1)}(x)$ (25) as follows:

$$N^2 = \int_0^1\left(\psi_n^{PT(1)}(x)\right)^2 r(x)\,dx.$$

<div align="right">(26)</div>

Then any limiting case $r(x) = \mathrm{const} \ne 1$ is covered automatically. Figures 34-47 demonstrate an integrally good accuracy of formula (25) normalized numerically with (26) for our relative big differences between $r_1$ and $r_2$.

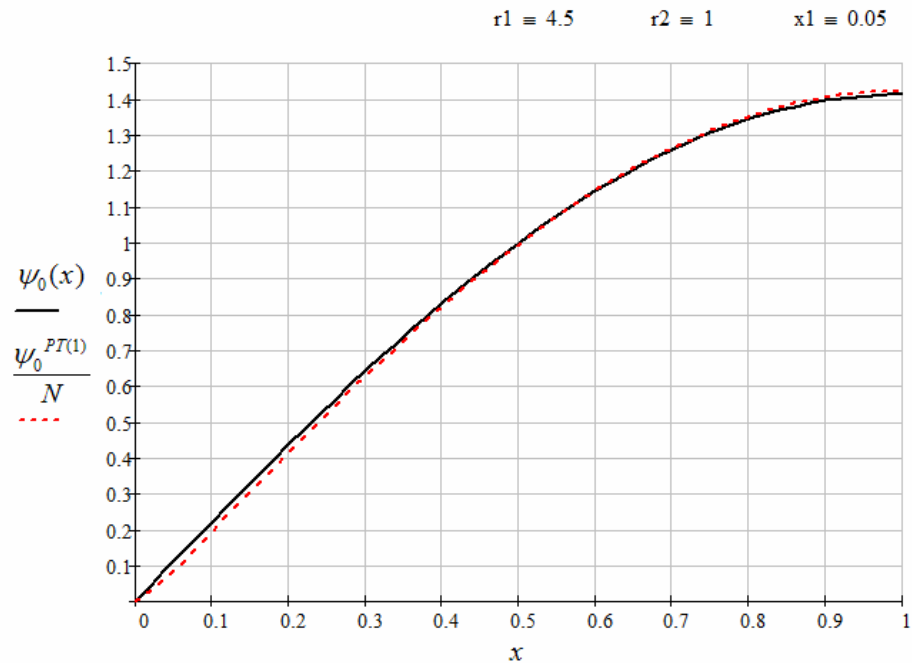

**Fig. 34.** Comparison of the first-order PT-approximation $\psi_0^{(1)}(x)$ with the exact eigenfunction.

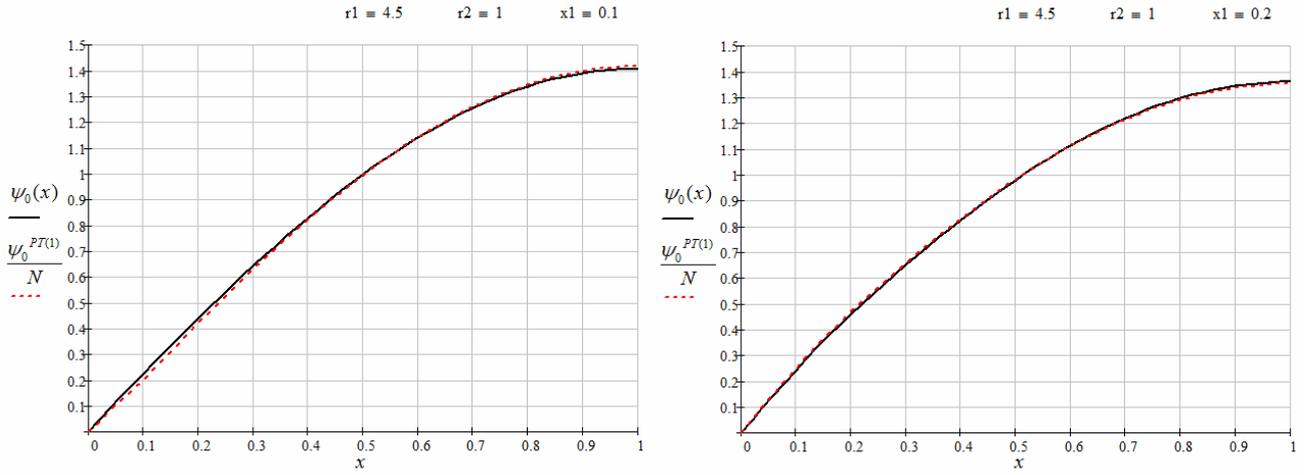

**Fig. 35.** Comparison of the first-order PT-approximation $\psi_0^{(1)}(x)$ with the exact eigenfunction.

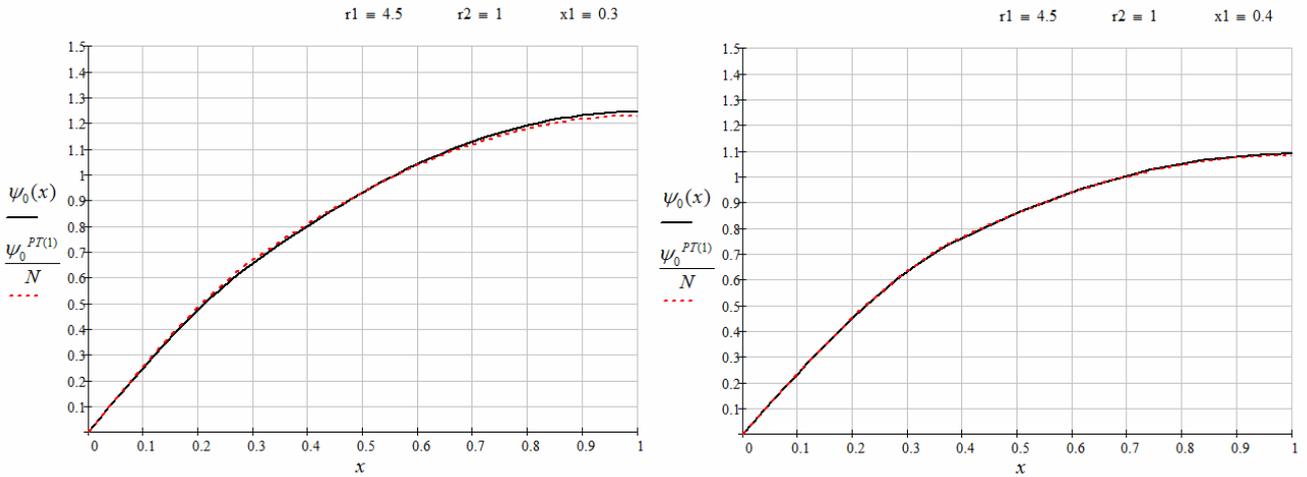

**Fig. 36.** Comparison of the first-order PT-approximation $\psi_0^{(1)}(x)$ with the exact eigenfunction.

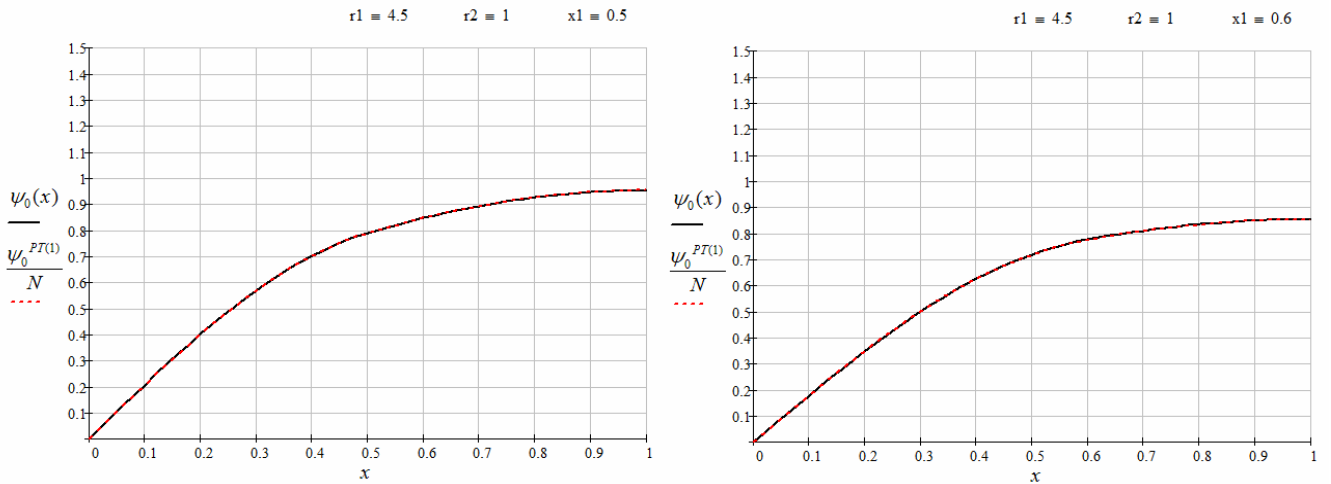

**Fig. 37.** Comparison of the first-order PT-approximation $\psi_0^{(1)}(x)$ with the exact eigenfunction.

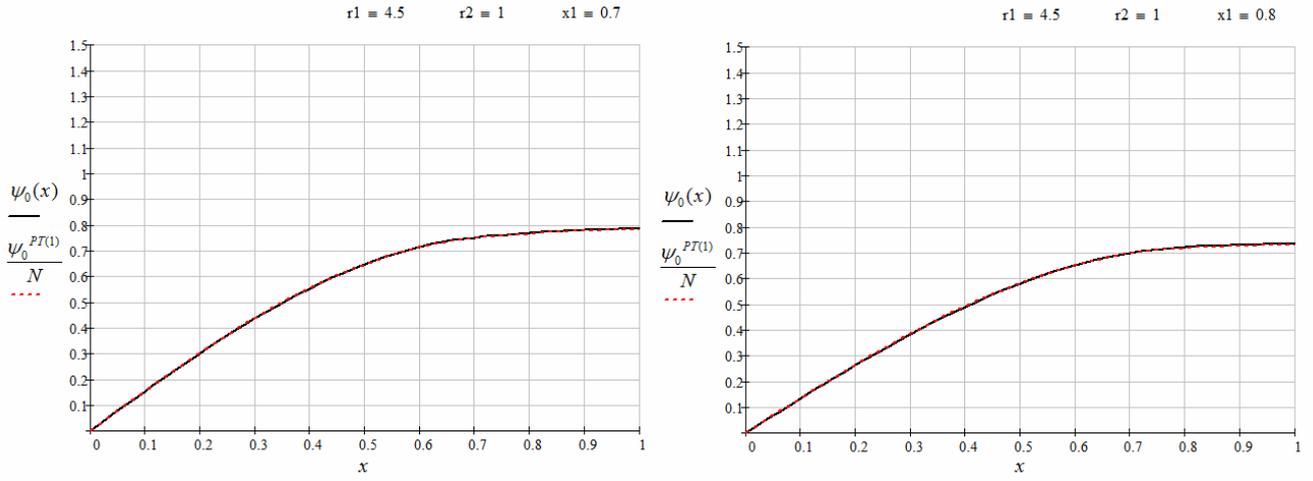

**Fig. 38.** Comparison of the first-order PT-approximation $\psi_0^{(1)}(x)$ with the exact eigenfunction.

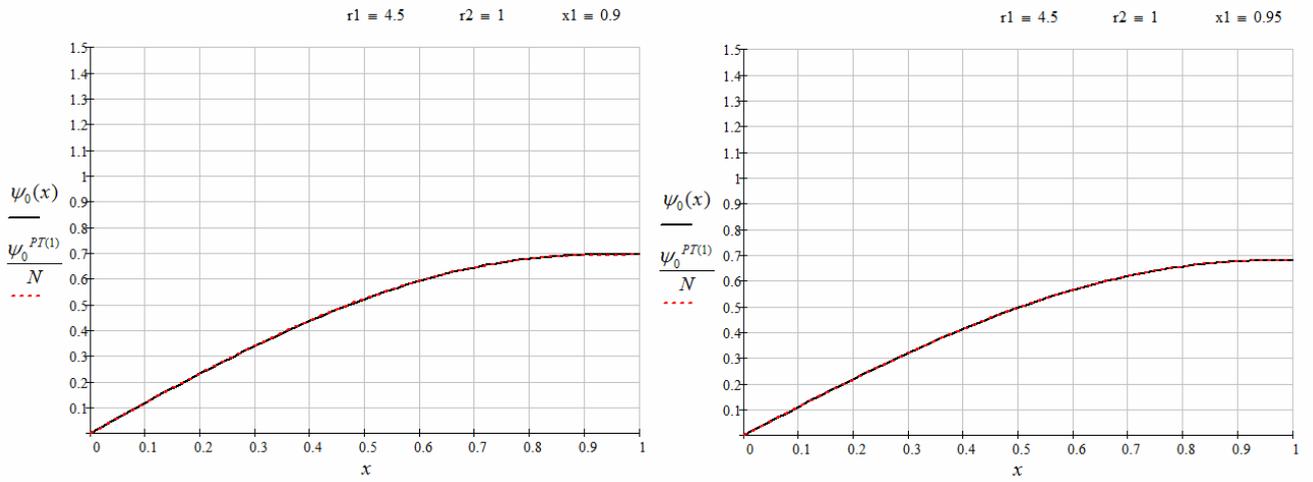

**Fig. 39.** Comparison of the first-order PT-approximation $\psi_0^{(1)}(x)$ with the exact eigenfunction.

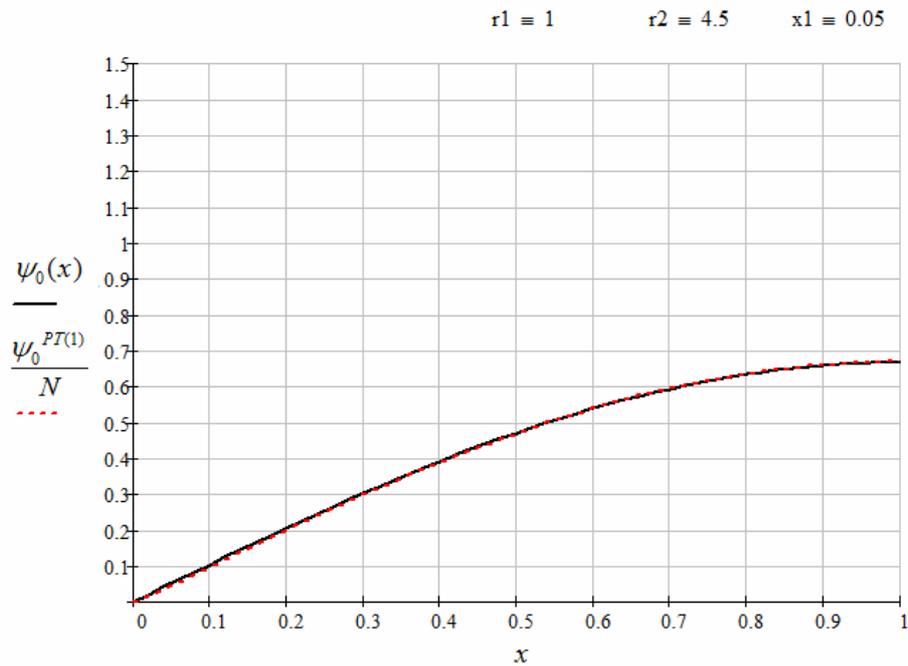

**Fig. 40.** Comparison of the first-order PT-approximation $\psi_0^{(1)}(x)$ with the exact eigenfunction.

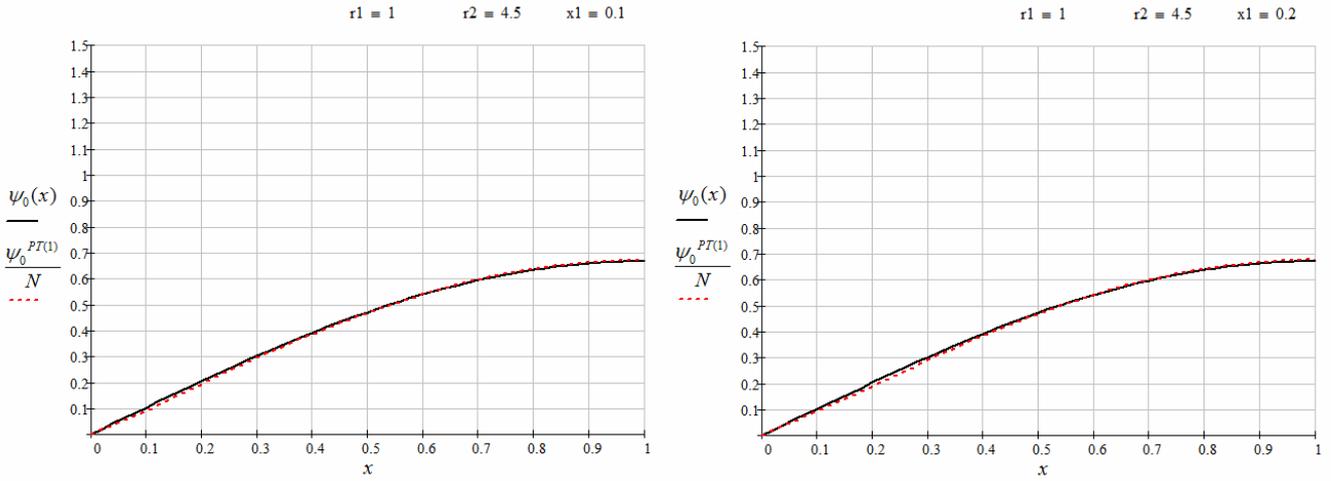

**Fig. 41.** Comparison of the first-order PT-approximation $\psi_0^{(1)}(x)$ with the exact eigenfunction.

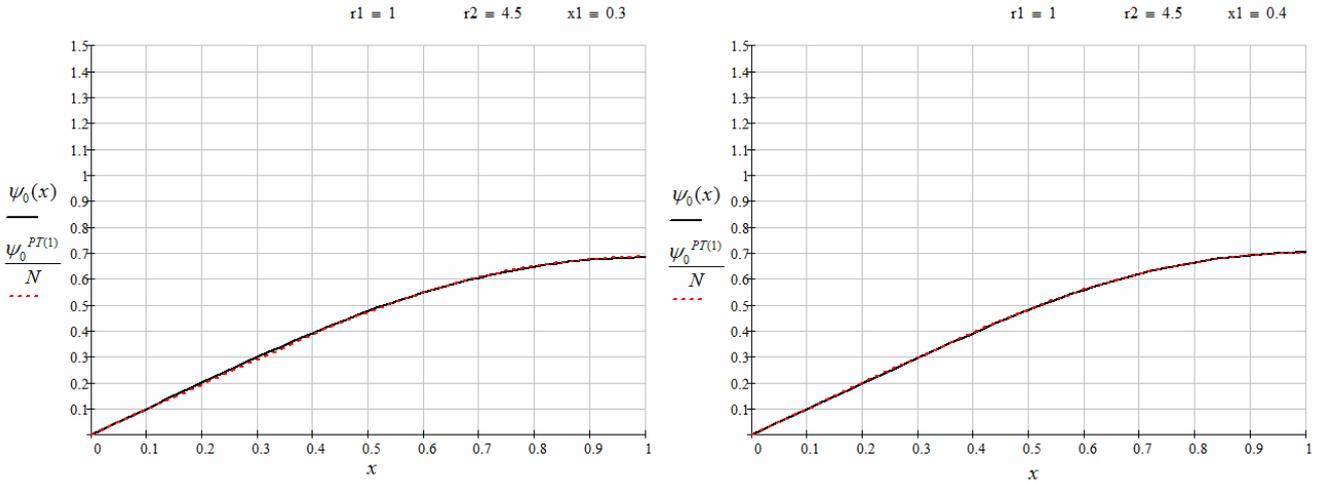

**Fig. 42.** Comparison of the first-order PT-approximation $\psi_0^{(1)}(x)$ with the exact eigenfunction.

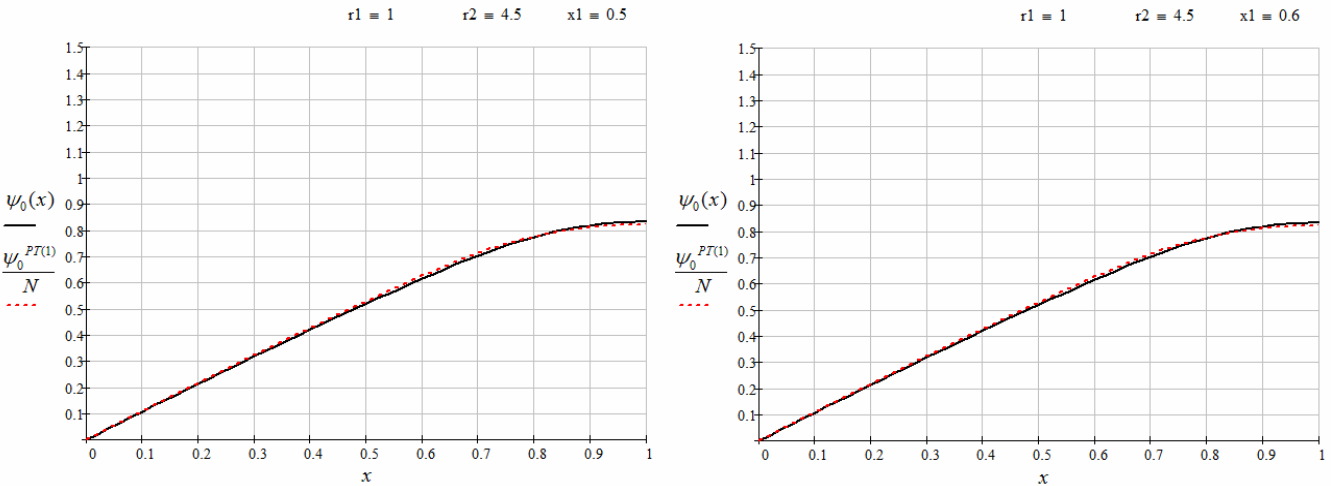

**Fig. 43.** Comparison of the first-order PT-approximation $\psi_0^{(1)}(x)$ with the exact eigenfunction.

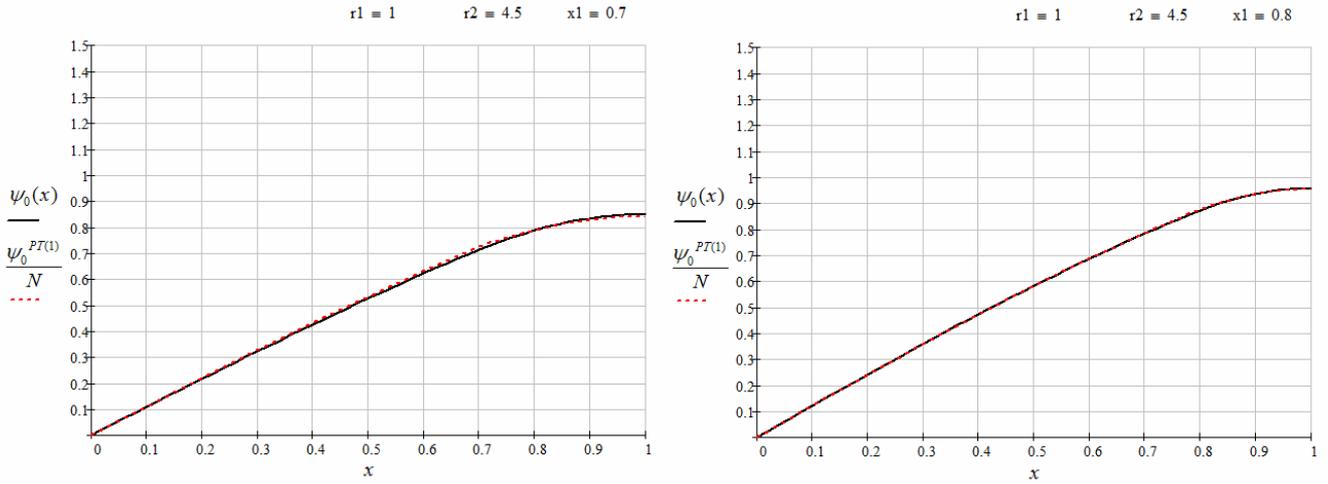

**Fig. 44.** Comparison of the first-order PT-approximation $\psi_0^{(1)}(x)$ with the exact eigenfunction.

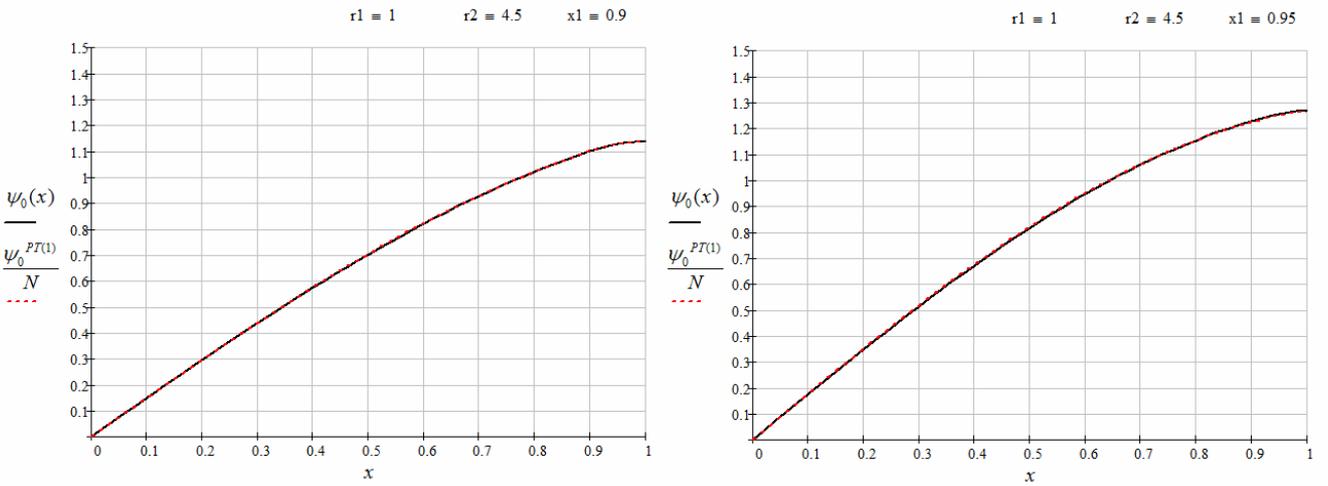

**Fig. 45.** Comparison of the first-order PT-approximation $\psi_0^{(1)}(x)$ with the exact eigenfunction.

Note, any zeroth-order eigenfunction $\psi_n^{(0)}(x) = \varphi_n^{(0)}(z(x))$ is continuous, but it changes its slope at $x = x_1$, like in Fig. 1 and in Fig. 46 (blue curves). The first-order correction, due to (24), eliminates essentially this slope change as well as practically removes the other zeroth-order solution inaccuracies (Fig. 46-49).

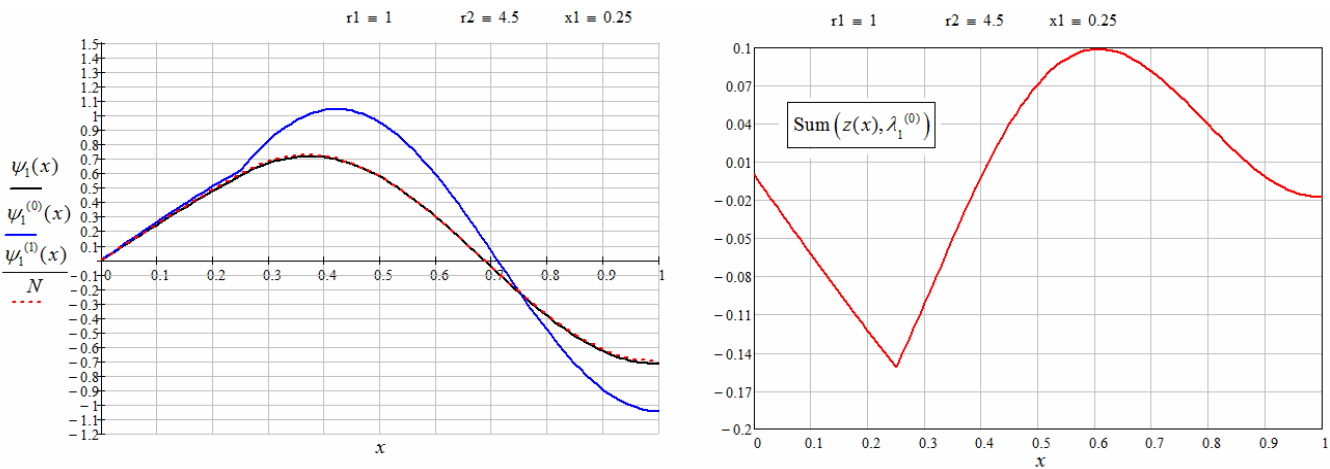

**Fig. 46.** Comparison of $\psi_1(x)$ with approximations $\psi_1^{(0)}(x)$ and $\psi_1^{(1)}(x)$ from (25)-(26).

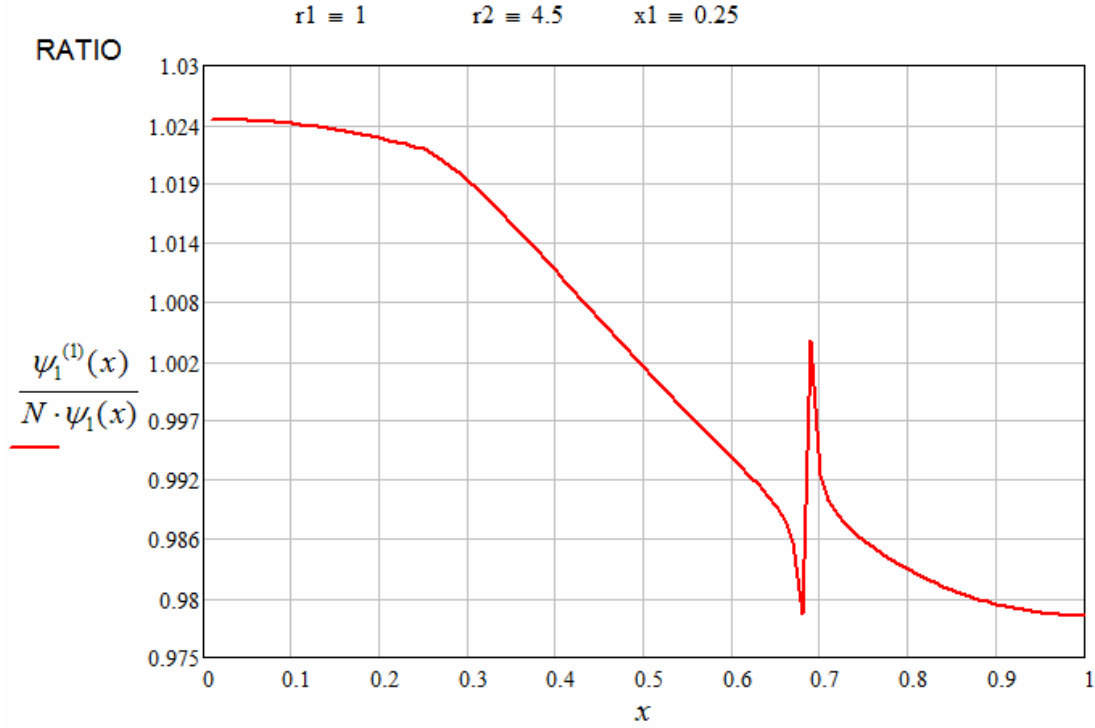

**Fig. 47.** Ratio of $\psi_1^{(1)}(x)$ from (25) with (26) and $\psi_1(x)$ for Fig. 46 is different from 1 by $\pm 2.5\%$ .

However locally the first order corrections may be "insufficient" when the "small parameter" $\ln\sqrt{r_1/r_2}$ is relatively big and the slope change is big too, like in Fig. 34, 50, 51. Then the higher-order PT-corrections may help. In this respect let us remember that the true "small parameter" of PT-expansion is not $\ln\sqrt{r_1/r_2}$, but $\xi = 2\left(\sqrt{r_1/r_2}-1\right)/\left(\sqrt{r_1/r_2}+1\right)$ [1], [2].

Another approach is to use the Green's function sum rules, as in **Sections 4-7 and 9**.

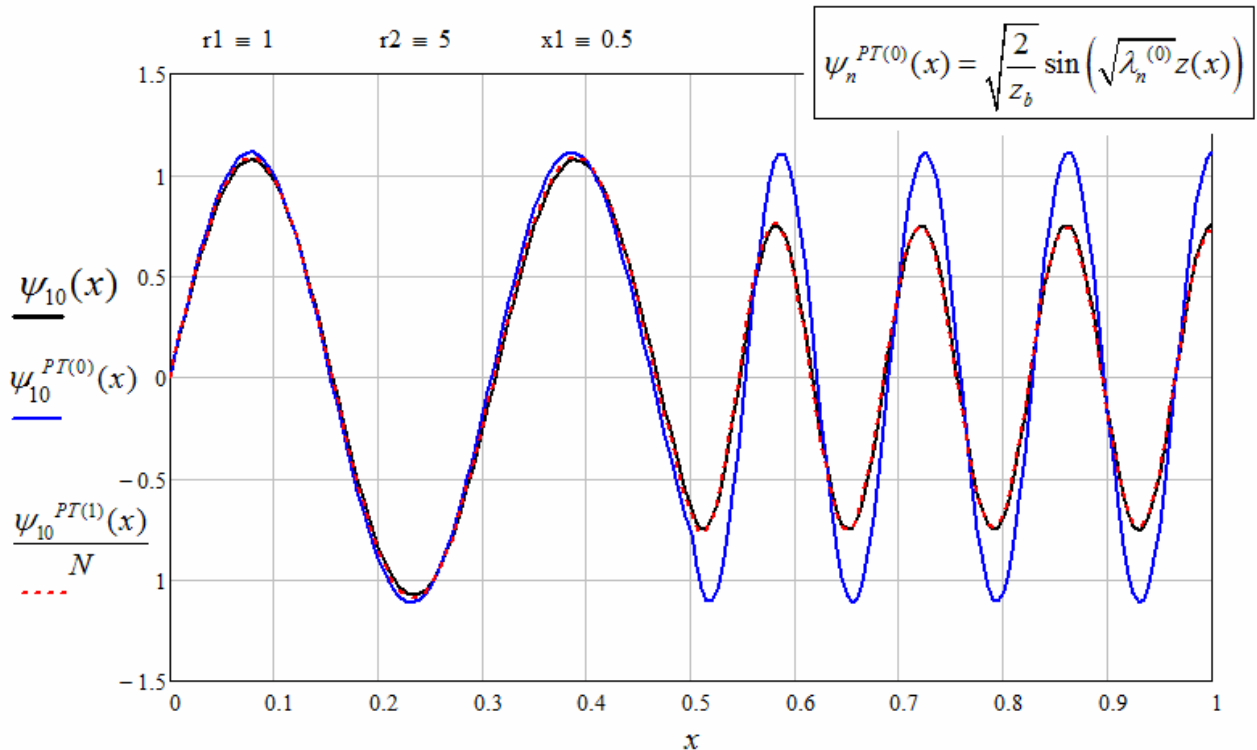

**Fig. 48.** Perturbation Theory results for the case shown in Fig. 1. $\psi_{10}^{PT(1)}(x)$ from (25) and $\psi_1(x)$ are normalized here.

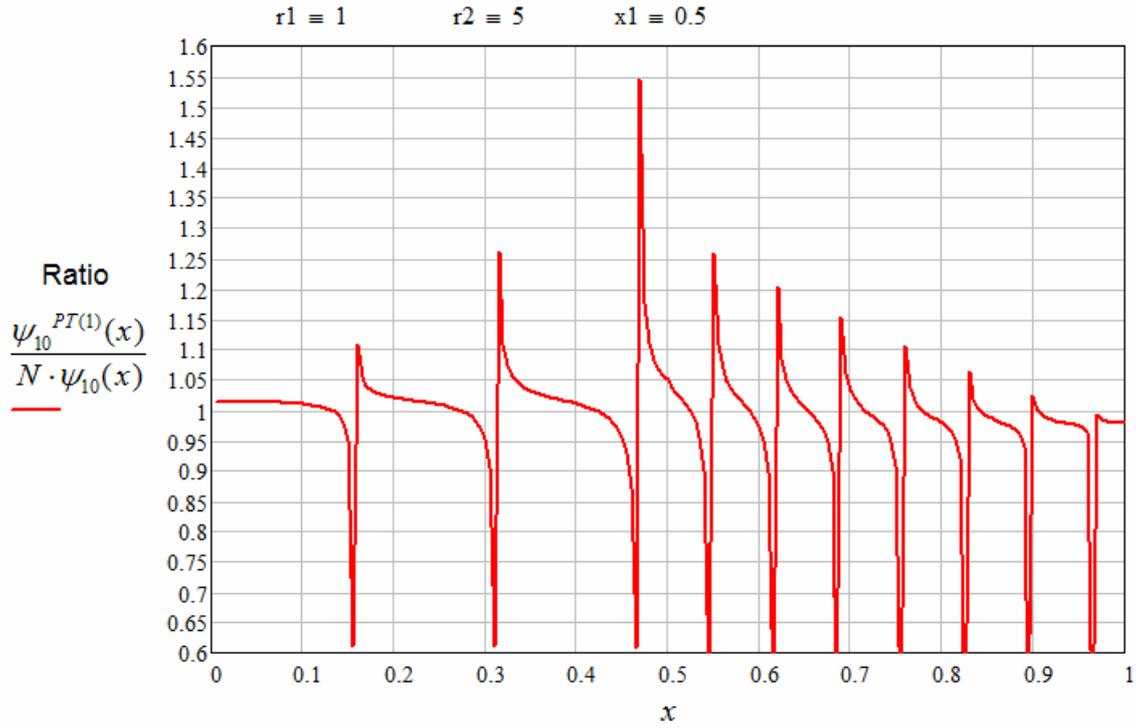

**Fig. 49.** Ratio of $\psi_{10}^{PT(1)}(x)$ from (25) with (26) and $\psi_1(x)$ for Fig. 48. Ratio only makes sense "far" from nodes.

## 9. Perturbation Theory results $\psi_{0,1}^{PT(1)}(x)$ as initial approximations for $\psi_{0,1}^{GF(1)}(x)$

In principle, the first-order PT-formulae (25)-(26) are also some analytical expressions of a descent accuracy and as such they can be used in Green's formulae (14) and (21) as initial approximations for iterative procedures instead of higher PT orders. I tested this approach too, see Fig. 50-53.

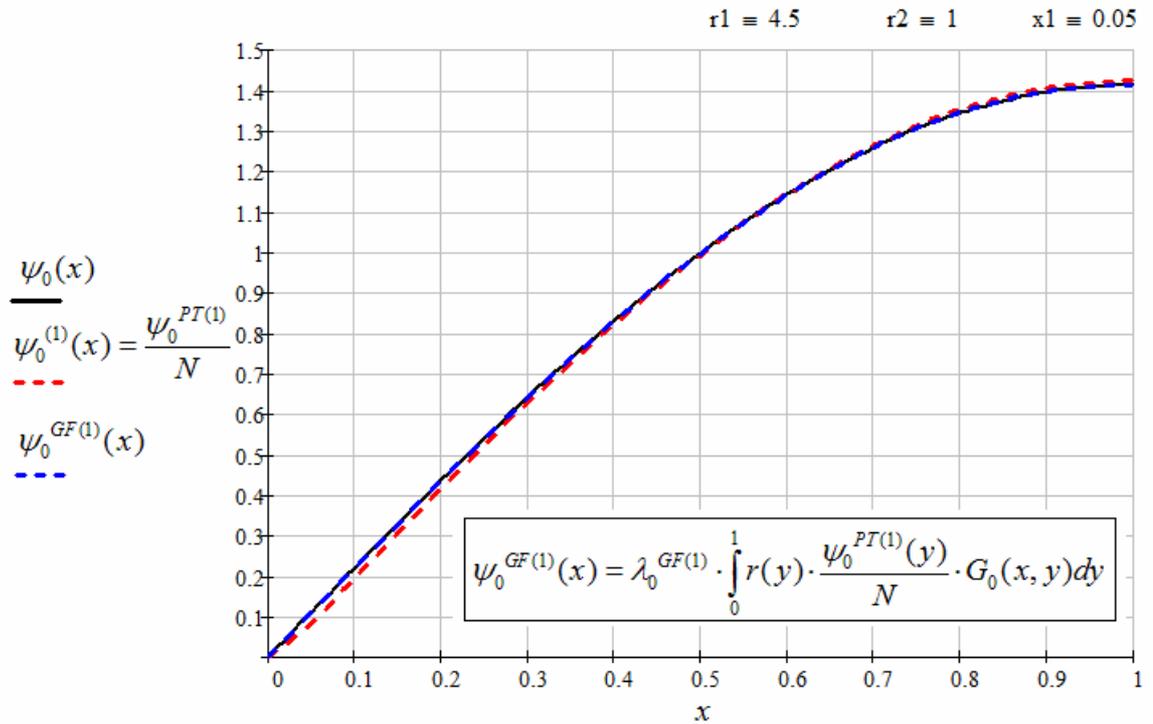

$$\psi_0^{GF(1)}(x) = \lambda_0^{GF(1)} \cdot \int_0^1 r(y) \cdot \frac{\psi_0^{PT(1)}(y)}{N} \cdot G_0(x,y) dy$$

**Fig. 50.** Case $n = 0$. Comparison of $\psi_0$ with $\psi_0^{PT(1)}$ from (25) with (26) and $\psi_0^{GF(1)}$ from (14) with use of $\psi_0^{PT(1)}$ on the right-hand side as an initial approximation. Radical improvement of local precision for a "difficult case" of a small thickness layer with a big $r_1$: $r_1 = 4.5$, $r_2 = 1$, $x_1 = 0.05$.

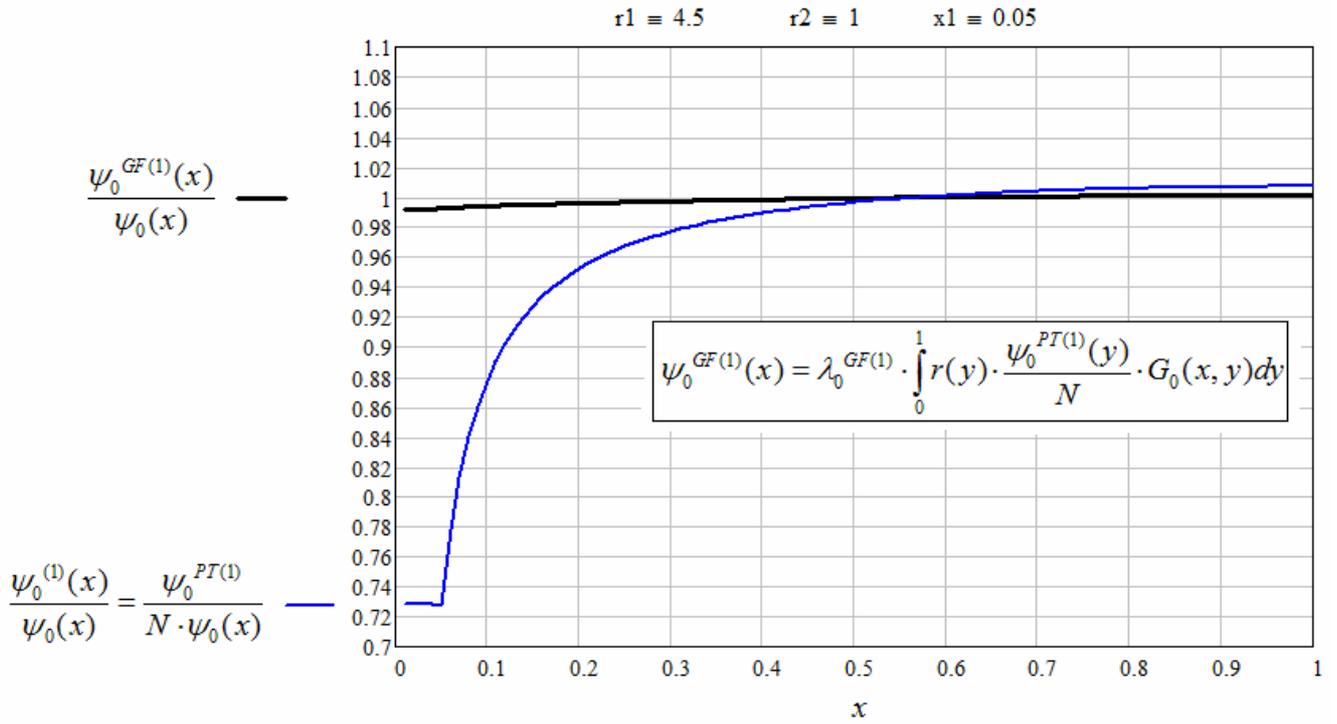

**Fig. 51.** Case $n = 0$. Ratios of $\psi_0^{PT(1)} / N$ and $\psi_0^{GF(1)}$ to the exact $\psi_0(x)$ for Fig. 50. Radical improvement of local precision for a "difficult case" of a small thickness layer with a big $r_1$: $r_1 = 4.5$, $r_2 = 1$, $x_1 = 0.05$.

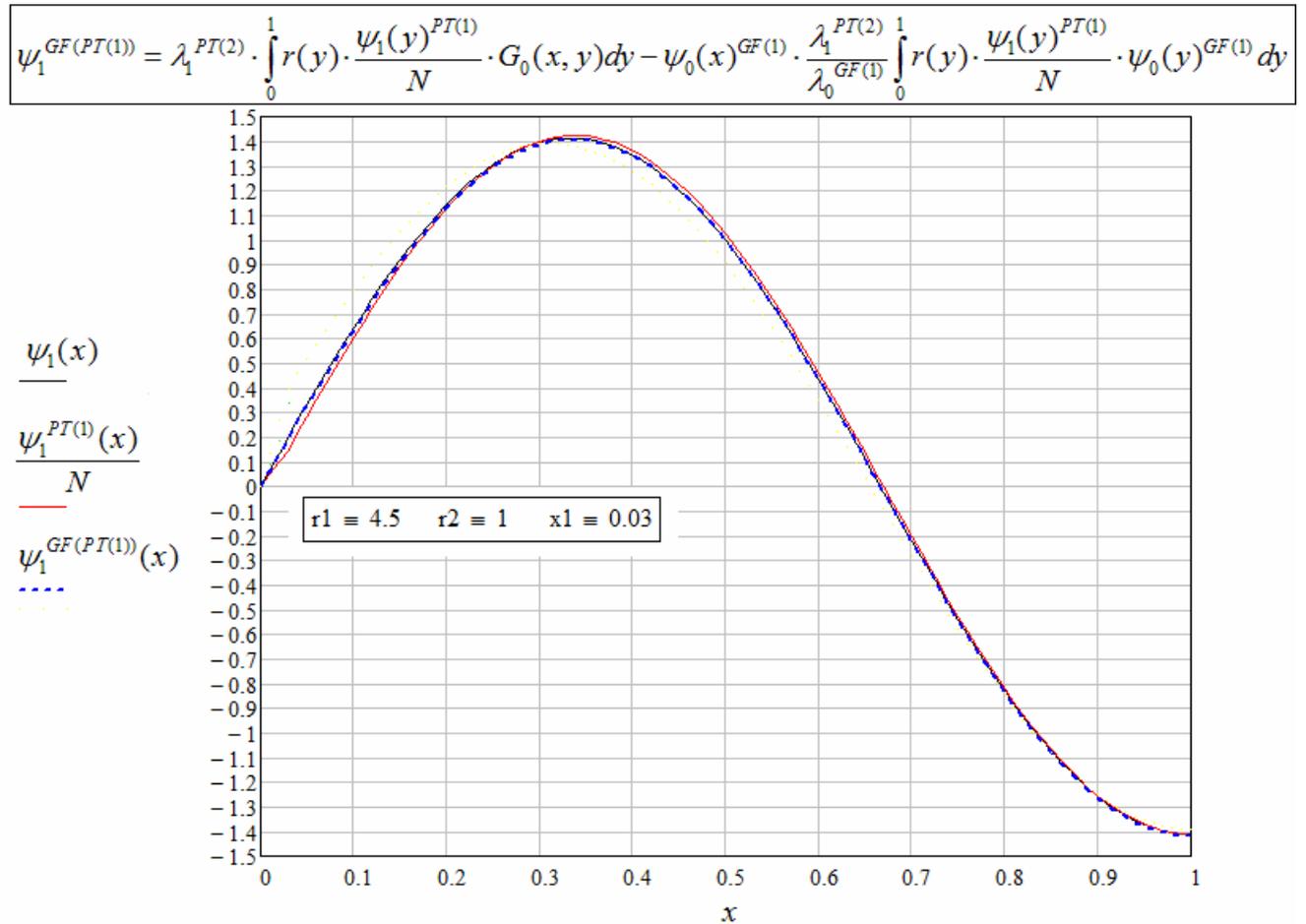

**Fig. 52.** Case $n = 1$. Comparison of $\psi_1$ with $\psi_1^{PT(1)}$ from (25) with (26) and $\psi_1^{GF(1)}$ from (14) with use of $\psi_1^{PT(1)}$ on the right-hand side as an initial approximation.

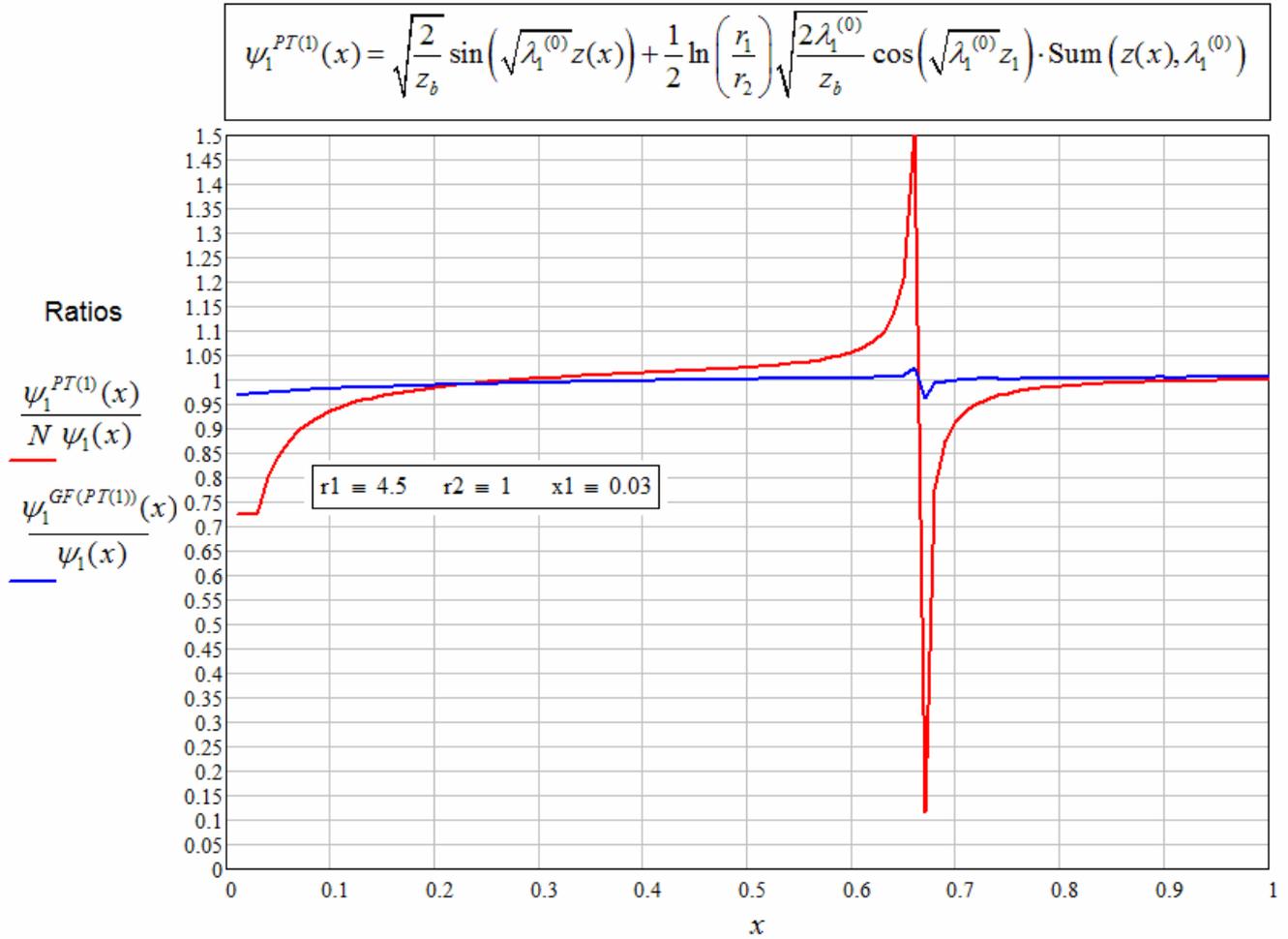

**Fig. 53.** Case $n = 1$. Ratios of $\psi_1^{PT(1)} / N$ and $\psi_0^{GF(1)}$ to the exact $\psi_0(x)$ for Fig. 52. Radical improvement of local precision for a "difficult case" of a small thickness layer with a big $r_1$: $r_1 = 4.5$, $r_2 = 1$, $x_1 = 0.03$.

For the other eigenfunctions $\psi_{n>2}(x)$ the first-order PT-formulae a la (25), (26) may be sufficient in practical applications due to their smaller contributions to the total solution $\Psi(x, t)$ of a non stationary problem.

## 10. Discussions

In the present paper I tested several analytical formulae for a specific Sturm-Liouville problem whose awkward perturbative formulation (**Section 1**) may have divergent matrix elements $U_{nm}$. Our better problem formulation (**Section 2**) follows just from good sense. As we could see, there in fact may be many different approximate analytical expressions for the same exact variable and **we are not bound to stick to the only one**, especially if the corresponding equation is badly guessed, like (3) for (1) with (4). Good analytical expressions are flexible and they help keep everything under control.

This paper deals with constructing **simple analytical approximations** for the solutions as it was conceived at the times of big and expensive computers with punched card input ("IBM machines") and small scientific calculators with simple programmed functions like $\sin(...)$, $\cos(...)$, $\sqrt{\phantom{x}}$, $x^y$, $\ln(...)$, etc., so "affordable" analytical constructions were of a great value due to simplifying qualitative analysis. I personally was intrigued by finding out the nature of the matrix element divergences in a problem having finite exact solutions and I did it [1], [2].

Nowadays many problems can be solved numerically with modern (super) computers, **but no computer can replace a creative researcher in correctly setting up a problem to solve**. For example, most "fundamental" QFT are still considered "ill-defined" or even "non existent" due to being "incomplete". The correct statement about them is that we still stick to a badly guessed QFT "by analogy with QED" with evident flaws and drawbacks. A better guess of equations and initial approximations may solve all those problems even in "incomplete" theories [4].

### 11. Conclusions

In the present paper I showed how important is to choose an appropriate initial approximation for building a reasonable perturbation theory or an iterative procedure. Although banal mathematically, this understanding is not widely appreciated in physics due to historical and some other (sociological) reasons. In particular, the renormalizations (modifications of bad solutions) have been given such a "state of the art" that it is extremely difficult to get through the common opinion today. In this article I demonstrate that one *can* find a **short-cut** to better equations with convergent series without appealing to "renormalizations". This short-cut consists generally in **reformulation** of the physical problems in better terms and variables, which better catch the main features of the exact solution qualitatively and thus quantitatively. The possibility of reformulating QFT in better terms should be taken seriously by the scientific community.

I believe that we can reformulate QED and some other "gauge" theories in order to directly obtain their final results without renormalizations.

### APPENDIX 1

As I mentioned in [1] (**Appendix 3**) the spectral sum $\sum_{m \neq n}^{\infty} \dfrac{\varphi_m^{(0)}(z) \varphi_m^{(0)\prime}(z_1)}{\lambda_n^{(0)} - \lambda_m^{(0)}}$ arising in the second-order correction to the eigenvalues is very sensitive to the spectral terms when $z$ is close to $z_1$, leaving apart that it is

slowly convergent and thus impractical for direct summation, see Fig. 54 plotted for $n = 0$:

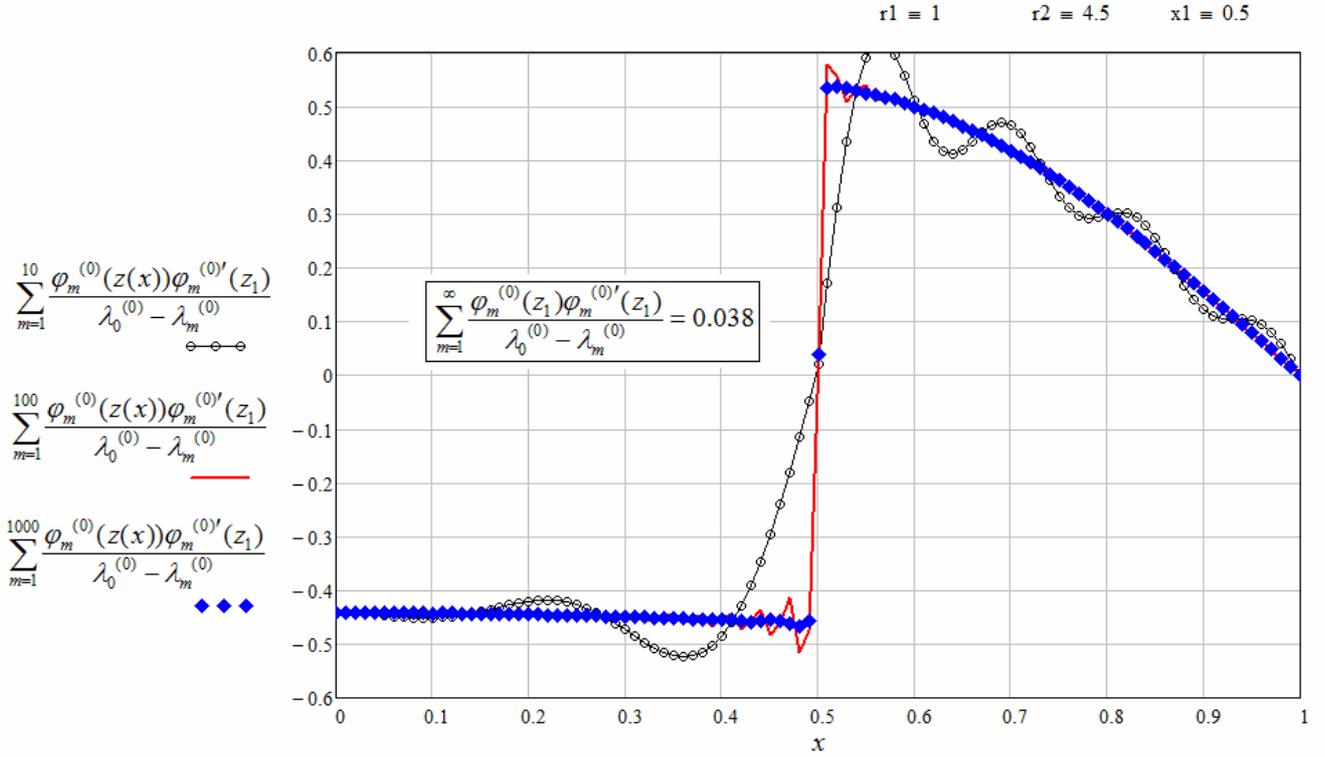

**Fig. 54.** Slow convergence and sensitivity of the spectral sum $\sum_{m \neq n}^{\infty} \dfrac{\varphi_m^{(0)}(z) \varphi_m^{(0)\prime}(z_1)}{\lambda_n^{(0)} - \lambda_m^{(0)}}$.

At $z = z_1$ this spectral sum is equal to:

$$\frac{1}{2}\left[ \sum_{m \neq n}^{\infty} \frac{\varphi_m^{(0)}(z_1 - 0) \varphi_m^{(0)\prime}(z_1)}{\lambda_n^{(0)} - \lambda_m^{(0)}} + \sum_{m \neq n}^{\infty} \frac{\varphi_m^{(0)}(z_1 + 0) \varphi_m^{(0)\prime}(z_1)}{\lambda_n^{(0)} - \lambda_m^{(0)}} \right], \tag{A1.1}$$

which can be calculated simply by differentiating (24b) on $z$ at $z = z_1$ and omitting the term arising from the derivative of the slope changing function $\sin\left( \sqrt{\lambda} \left| z_1 - z \right| \right)$. The resulting analytical formula is short and simple:

$$\frac{1}{2} \left\{ \frac{\sin\left( 2\sqrt{\lambda} z_1 \right)}{2\sqrt{\lambda} z_b} - \left( 1 - 2 \frac{z_1}{z_b} \right) \cos\left( 2\sqrt{\lambda} z_1 \right) \right\}. \tag{A1.2}$$

From this one easily obtains the second-order PT-approximation given in the main text:

$$\lambda_n^{PT(2)} = \lambda_n^{PT(1)} + \varepsilon^2 \frac{\sqrt{\lambda_n^{(0)}} \sin\left( 2\sqrt{\lambda_n^{(0)}} z_1 \right)}{8 z_b} \left\{ \frac{\sin\left( 2\sqrt{\lambda_n^{(0)}} z_1 \right)}{2\sqrt{\lambda_n^{(0)}} z_b} - \left( 1 - 2 \frac{z_1}{z_b} \right) \cos\left( 2\sqrt{\lambda_n^{(0)}} z_1 \right) \right\}. \tag{A1.3}$$

### APPENDIX 2

In [5] it was proposed to use some auxiliary functions for improving precision, for example:

$$IGF(x, y) = \int_0^1 G_0(x, x') G_0(x', y) r(x') dx',\qquad (A2.1)$$

whose spectral representation contains the **squares** of eigenvalues in denominators and is the following:

$$IGF(x, y) = \sum_{m=0}^{\infty} \frac{\psi_m(x)\psi_m(y)}{\lambda_m{}^2}\qquad (A2.2)$$

It means reinforcing suppression factors in formulae like (14), (20) discussed above, and thus increase of precision of calculated eigenfunctions (the projections $(\psi_{0,1}{}^{(0)}, \psi_m)$ remain, of course, the same).

The authors of [5] awkwardly call such auxiliary functions with multiple integrations "iterated Green's functions $G_n(x, y)$" where "$n$" in [5] stands for the number of iteration. Such constructions arise naturally while iterating the equations (12), and (18) for finding $\psi_{0,1}$ (see, for example, (A2.5)), but $G_{n\to\infty}$ does not converge to any exact Green's function, for example, to (9), so I do not use these misleading notations and names from [5]. I call these auxiliary functions the "*integrated* Green's functions" (IGF) in the sense (A2.1). In fact, such functions are not Green's functions at all; they obey the equations like:

$$\frac{d^2}{dx^2} IGF(x, y) = -G_0(x, y) r(x), \quad IGF(0, y) = 0, \quad IGF'(1, y) = 0,\qquad (A2.3)$$

i.e., they obey ordinary differential equations with no $\delta$-function on the right-hand side; (A2.3) is similar to equation (12), but the "source" term in (A2.3) is a known function containing $y$ as a parameter. Formally such equations can be solved via double integration and applying the boundary conditions. Practically, such a double integration may be rather laborious. I tested this approach for our two-layer problem (1) with (4). The "free" Green's function may be represented simply as follows: $G_0(x, y) = \min(x, y)$. The analytical solution to (A2.3) can be obtained directly from definition (A2.1) with help of one integration:

$$G_0(x, x') = x\theta(x' - x) + x'\theta(x - x') = \min(x, x'), \quad r(x) = r_1\theta(x_1 - x) + r_2\theta(x - x_1),$$

$$IGF(x, y) = \int_0^1 \left[x\theta(x' - x) + x'\theta(x - x')\right]\left[x'\theta(y - x') + y\theta(x' - y)\right]\left[r_1\theta(x_1 - x') + r_2\theta(x' - x_1)\right] dx',$$

or by double integration of (A2.3). In any case one may obtain an analytical formula like this one:

$$IFG(x, y) = x\left[\frac{r_1 - r_2}{2}\left(\min(y, x_1) - y\right)^2 + y\left((r_1 - r_2)x_1 - r_1\frac{y}{2} + r_2\right) - r_1\frac{x^2}{6}\right] -$$
$$\frac{r_1 - r_2}{2}\left\{\left[\left(x + y - \frac{2}{3}\min(y, x_1)\right)\min(y, x_1) - 2xy\right]\min(y, x_1) - \left(\frac{x}{3} - y\right)x^2\right\}\theta(x - \min(y, x_1)) -$$
$$r_2\frac{(y - x)^3}{6}\theta(x - y) + \frac{y(x - x_1)^2}{2}(r_1 - r_2)\theta(x - x_1).\qquad (A2.4)$$

This is a longer formula, with no integration anymore, but with several switching functions $\theta(...)$ and $\min(...)$, and further *analytical* integrations with it are even more laborious, although still possible. In fact, we have already implicitly used $IGF(x, y)$ when we were calculating the next iteration in **Section 6**. Indeed, formally

$$\psi_0(x) = \lambda_0 \int_0^1 r(y) \cdot \psi_0(y) \cdot G_0(x, y) dy \Rightarrow \psi_0{}^{GF(0)}(x) = \lambda_0 \int_0^1 r(y) \cdot \psi_0{}^{(0)}(y) \cdot G_0(x, y) dy,$$

$$\psi_0{}^{GF(1)}(x) = \lambda_0 \int_0^1 r(y) \cdot \psi_0{}^{GF(0)}(y) \cdot G_0(x,y) dy = \lambda_0{}^2 \int_0^1 r(y) \cdot \psi_0{}^{(0)}(y) \cdot IGF(x,y) dy \qquad (A2.5)$$

Unlike $G_0(x,y)$, function $IGF(x,y)$ depends on $r(x)$ by definition. Below there is a picture of the auxiliary function calculated numerically (A2.1) and analytically (A2.4) for some fixed values of $y$, $r_1$, $r_2$, $x_1$:

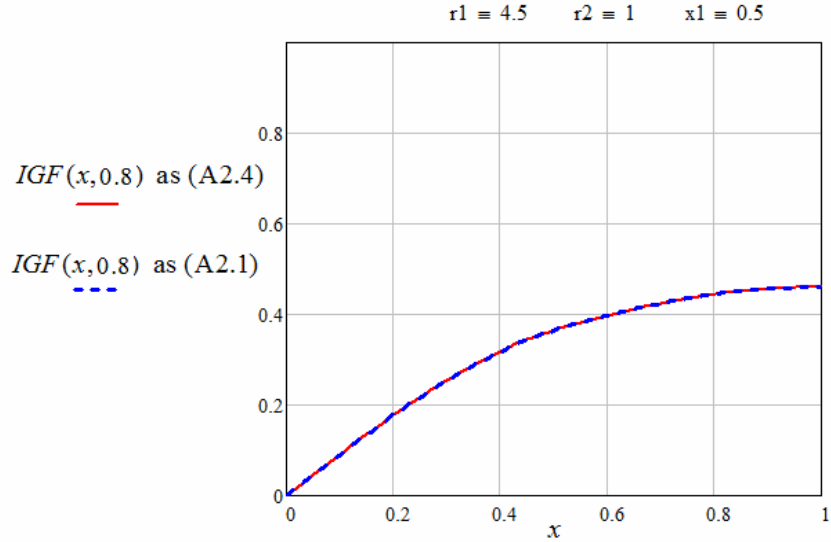

**Fig. 55.** Auxiliary function $IGF(x, 0.8)$ for particular values of input parameters.

As we could see in the main text, we obtained nicely convergent iterations in case $n = 0$ (**Section 6**), but we encountered badly convergent results in case $n = 1$ for our relatively big differences between $r_1$ and $r_2$. A formal use of (A2.5) for $n = 1$ does not work either, and for "extracting" a good $\psi_1(x)$ we must subtract an "amplified" bad contribution of projection $\left(\psi_1{}^{(0)}, \psi_0\right)$, as it has been done in **Section 7** in formulae (19)-(21):

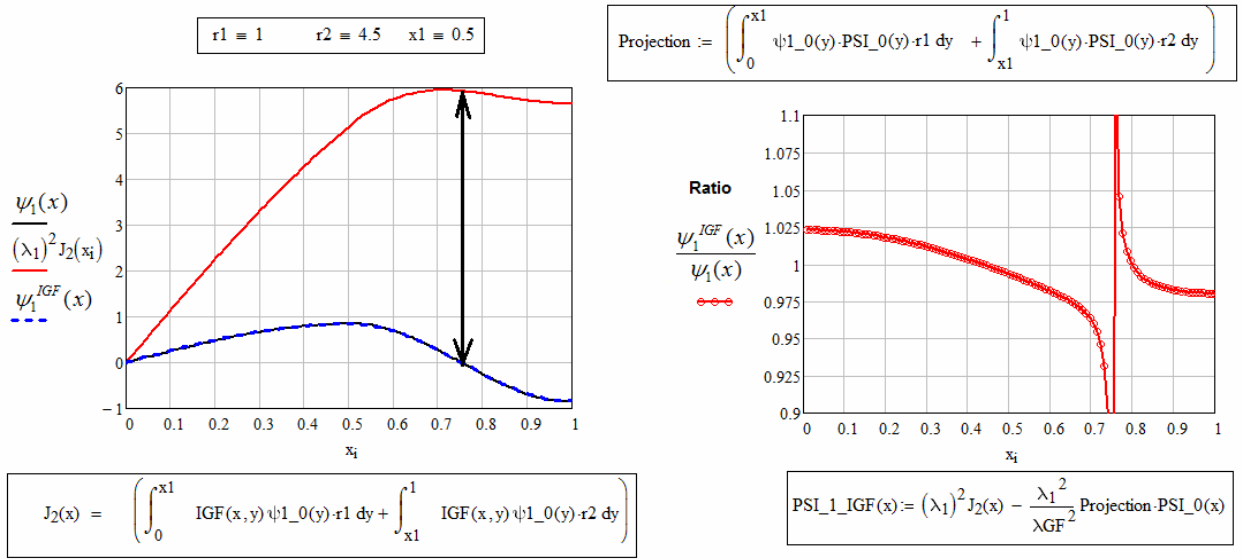

**Fig. 56.** Reformulation of the iterative procedure for calculating $\psi_1$ with help of $IGF(x, y)$: the subtraction of projection is an absolutely necessary element of calculation (compare it with Fig. 27). Here $\mathrm{PSI\_0}(x) \equiv \psi_0(x)$ (exact eigenfunction), $\mathrm{PSI\_1\_IGF}(x) \equiv \psi_1{}^{IGF}(x)$, $\psi1\_0(x) \equiv \psi_1{}^{(0)}(x) = \varphi_1{}^{(0)}(z(x)) \, r(x)^{-1/4}$ (approximate ones), $\lambda\mathrm{GF} \equiv \lambda_0{}^{GF(1)}$, $r1 \equiv r_1$, $r2 \equiv r_2$, $x1 \equiv x_1$, $x_i \equiv i/100$, $i = 0,1,...,100$. Each integral from $0$ to $1$ is split in two integrals with constant $r(x)$ in order to speed up numerical integration.

Analytical formulae like (A2.4) and sum rules (A2.2) can be used not only with WKB-like initial approximations, but also with PT-solutions (25), (26) for more precise calculations.

**APPENDIX 3**

### A3.1. A formal proof of convergence of iterative procedure for $\psi_0(x)$

Let us represent the initial approximation $\psi_0^{(0)}(x)$, whatever it is, as a superposition of the exact, but unknown, eigenfunctions:

$$\psi_0(x)^{(0)} = a_0\psi_0(x) + \sum_{n=1}^{\infty} a_n\psi_n(x), \ a_0 \approx 1, \ \left|a_{n>0}\right| \ll 1. \ \text{Then} \tag{A3.1}$$

$$\psi_0^{GF(0)}(x) \equiv I_1(x) = \sum_{n=0}^{\infty} \psi_n(x) \frac{\left(\psi_0(y)^{(0)}, \psi_n(y)\right)}{\lambda_n} = \frac{a_0}{\lambda_0}\left(\psi_0(x) + \sum_{n=1}^{\infty} \psi_n(x)\frac{a_n}{a_0}\frac{\lambda_0}{\lambda_n}\right), \tag{A3.2}$$

$$\psi_0^{GF(1)}(x) \equiv I_2(x) = \sum_{n=0}^{\infty} \psi_n(x) \frac{\left(\psi_0^{GF(0)}(y), \psi_n(y)\right)}{\lambda_n} = \frac{a_0}{\lambda_0^{2}}\left[\psi_0(x) + \sum_{n=1}^{\infty} \psi_n(x)\frac{a_n}{a_0}\left(\frac{\lambda_0}{\lambda_n}\right)^2\right], \tag{A3.3}$$

etc. The ground state eigenfunction is equal exactly to $\psi_0(x) = I_i(x)\dfrac{\lambda_0^{\,i}}{a_0} - \sum_{n=1}^{\infty}\psi_n(x)\dfrac{a_n}{a_0}\left(\dfrac{\lambda_0}{\lambda_n}\right)^i$, $i = 1, 2, \ldots$ and approximately to $\psi_0^{GF(i-1)}(x) = I_i(x)\left(\lambda_0^{\,i}/a_0\right)$ due to suppression factors $\left|a_n/a_0\right| \ll 1$ and $\left(\lambda_0/\lambda_n\right)^i \ll 1$ in the remaining sum. Decreasing coefficients $\left(a_n\right)^{(i)} \equiv a_n\left(\lambda_0/\lambda_n\right)^i = \left(a_n\right)^{(i-1)}\left(\lambda_0/\lambda_n\right)$ reflect "improved orthogonality" of $\psi_0^{GF(i)}(x)$ to $\psi_{n\geq 1}$ in (A3.1) in comparison with orthogonality of $\psi_0^{GF(i-1)}(x)$ used as an initial approximation for calculating $\psi_0^{GF(i)}(x)$, as it was mentioned in **Section 6**.

As the suppression factors $\left(\lambda_0/\lambda_n\right)^i$ decrease when the iteration number $i$ grows, the iterative procedure converges to $a_0 \cdot \psi_0/\lambda_0^{\,i}$. In reality we are not supposed to know the coefficient $a_0$ nor $\lambda_0$, hence the approximate functions $\psi_0^{GF(0)}(x), \psi_0^{GF(1)}(x)$ from (A3.2), (A3.3), etc., need in principle an additional normalization procedure.

### A3.2 A formal proof of convergence of iterative procedure for $\psi_1(x)$

Let us represent the initial approximation $\psi_1^{(0)}(x)$, whatever it is, as a superposition of the exact eigenfunctions:

$$\psi_1(x)^{(0)} = b_0\psi_0(x) + b_1\psi_1(x) + \sum_{n=2}^{\infty} b_n\psi_n(x), \ b_1 \approx 1, \ \left|b_{n\neq 1}\right| \ll 1. \qquad \text{Then} \tag{A3.4}$$

$$J_1(x) = \sum_{n=0}^{\infty} \psi_n(x)\frac{\left(\psi_1(y)^{(0)}, \psi_n(y)\right)}{\lambda_n} = \psi_0(x)\frac{b_0}{\lambda_0} + \psi_1(x)\frac{b_1}{\lambda_1} + \sum_{n=2}^{\infty} \psi_n(x)\frac{b_n}{\lambda_n}, \ldots \tag{A3.5}$$

$$J_i(x) = \sum_{n=0}^{\infty} \psi_n(x)\frac{\left(\psi_1(y)^{(0)}, \psi_n(y)\right)}{\lambda_n^{\,i}} = \psi_0(x)\frac{b_0}{\lambda_0^{\,i}} + \psi_1(x)\frac{b_1}{\lambda_1^{\,i}} + \sum_{n=2}^{\infty} \psi_n(x)\frac{b_n}{\lambda_n^{\,i}}, \tag{A3.6}$$

$$\psi_1(x) = J_i(x)\frac{\lambda_1^{\,i}}{b_1} - \psi_0(x)\frac{b_0}{b_1}\left(\frac{\lambda_1}{\lambda_0}\right)^i - \sum_{n=2}^{\infty}\psi_n\frac{b_n}{b_1}\left(\frac{\lambda_1}{\lambda_n}\right)^i \Rightarrow \psi_1^{GF}(x) = J_i(x)\frac{\lambda_1^{\,i}}{b_1} - \psi_0(x)\frac{b_0}{b_1}\left(\frac{\lambda_1}{\lambda_0}\right)^i \tag{A3.7}$$

due to suppression factors $\left|b_{n\geq 2}/b_1\right| \ll 1$ and $\left(\lambda_1/\lambda_{n\geq 2}\right)^i \ll 1$. As the factor $\left(\lambda_1/\lambda_0\right)^i$ at $\psi_0(x)$ is big and only grows when $i$ increases, the "parasite contribution" of $\psi_0(x)$ in (A3.7) is left subtracted like in (20), (21). Otherwise the iterative procedure (18) progressively diverges with $i$.

Let us note that for initially very small projection $|\delta_0| \ll 1$ the first iteration of (18) may "work", i.e., it may improve the initially bad precision without subtraction because of essential suppression of contributions of $\psi_{n>1}(x)$ in the initial approximation $\psi_1(x)^{(0)}$ (see an example in **Fig. 57-59**), and such a calculation is simpler. However each next iteration multiplies the admixture of the ground state $\psi_0(x)$ in $J_i(x)$ by the factor $(\lambda_1 / \lambda_0) \gg 1$, so subtraction is generally needed for dealing in practice with a stable and convergent iterative procedure, as it was done in **Section 7** and in **Fig. 56**.

(Theoretically, we can advance a two-subtraction scheme for calculating $\psi_2(x)$, why not, but with use of $IFG(x, y)$ (A2.4) and PT-solutions (25) for a better accuracy because simple suppression factors $(\lambda_2 / \lambda_{n>2}) = 25 / 49 \approx 0.51, \ 25 / 81 \approx 0.31, \dots$ are not very small now and the contribution of discarded sums may be important (it depends, of course, on quality of the initial approximation $\psi_2^{(0)}$ ).)

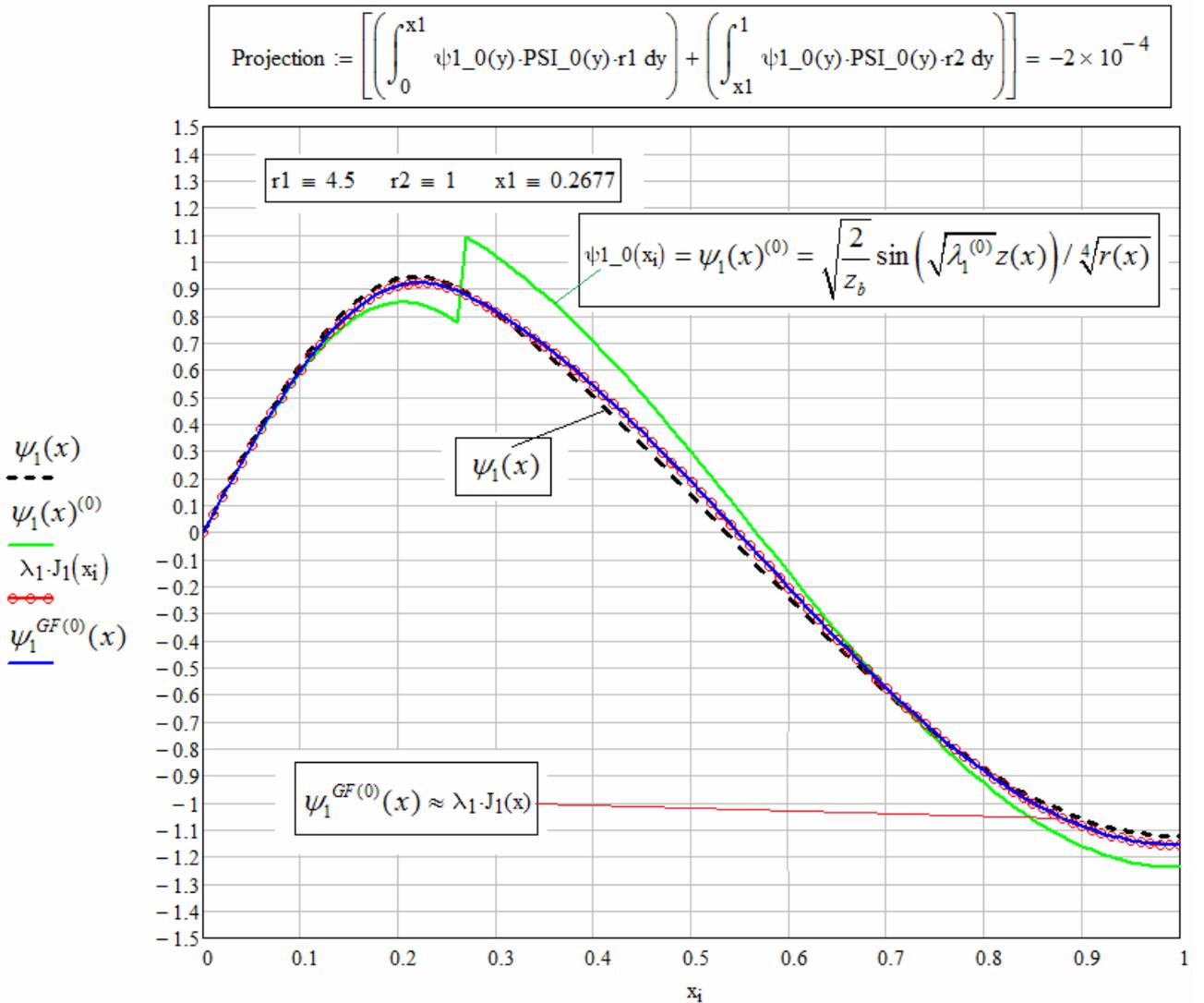

**Fig. 57.** Here $x_1$ was chosen so that a small value of projection is obtained (the blue and the red lines coincide). Then formula (18) works without subtraction due to too small contribution of $\psi_0$ into $\psi_1^{(0)}$: $\psi_1^{GF(0)}$ is better than $\psi_1^{(0)}$, but this is an exotic case. Normally one needs a convergent iteration procedure to advance without fail (compare Fig. 57 with Fig. 19). The visible difference between $\psi_1$ and $\psi_1^{GF(0)}$ in this figure ($\approx \pm 3\%$, see Fig. 59) is exclusively due to neglecting the sum $\sum_{n=2}^{\infty} \psi_n(x)(b_n / b_1)(\lambda_1 / \lambda_n)$. Here $\mathrm{PSI\_1}(x) \equiv \psi_1(x)$, $\mathrm{PSI\_1\_GF}(x) \equiv \psi_1^{GF(0)}(x) \approx \lambda_1 \cdot J_1(x)$, $\psi1\_0(x) \equiv \psi_1^{(0)}(x) = \varphi_1^{(0)}(z(x)) / \sqrt[4]{r(x)}$ (a WKB-like one). No additional normalization was applied.

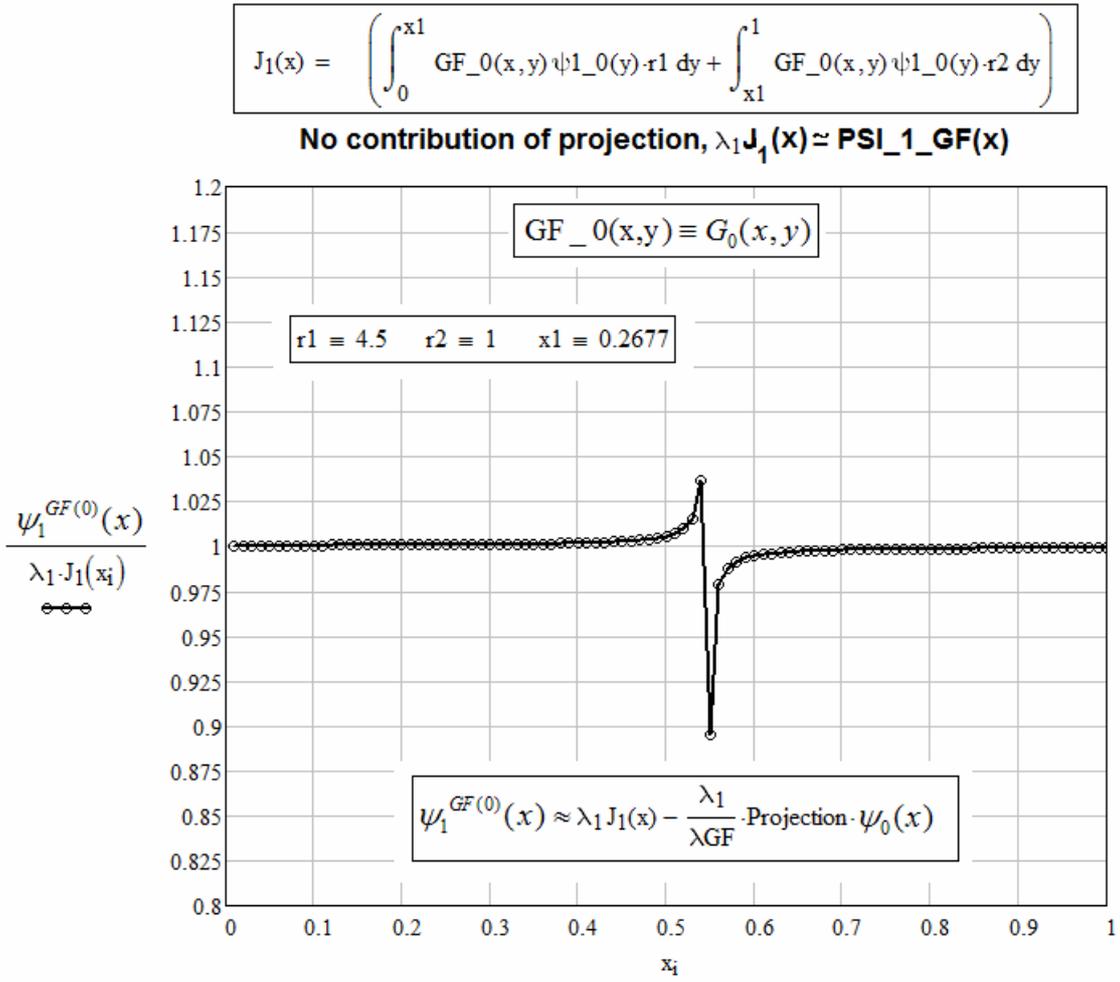

**Fig. 58.** In this exotic case (Fig. 57) formula (18) gives nearly the same results as formula (20).

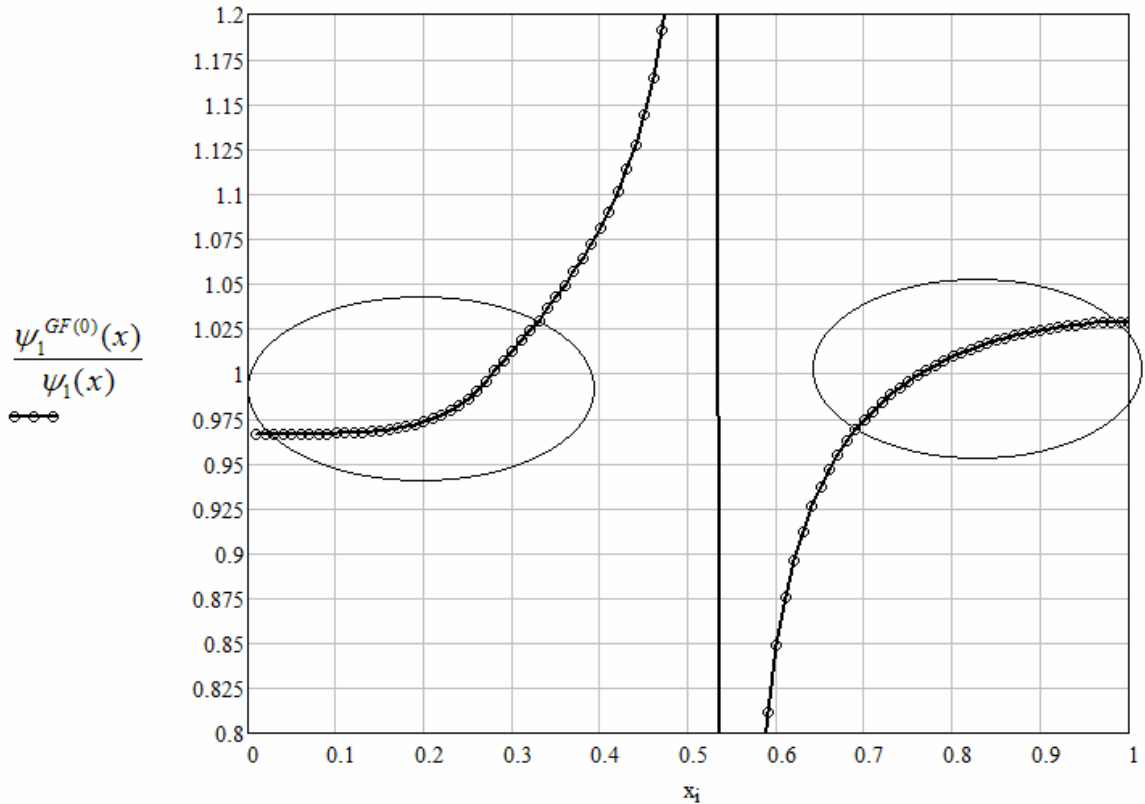

**Fig. 59.** In this exotic case (Fig. 57) applying formula (18) improves the precision with respect to that of $\psi_1^{(0)}(x)$.